\documentclass[12pt, preprint]{aastex}
\usepackage{amsmath}
\usepackage{graphicx}
\usepackage{natbib}

\usepackage[colorlinks=true,citecolor=blue,breaklinks=true,linktocpage=true]{hyperref}
\bibpunct{(}{)}{;}{a}{}{,} 
\usepackage[pagewise]{lineno}
\usepackage{xspace}
\usepackage{hyperref}

\usepackage{multirow}

\def\vec#1{\ensuremath{\mathbf{#1}}}

\newcommand{\kc}{\ensuremath{k_{\rm c}}}
\newcommand{\kcl}{\ensuremath{k_{{\rm c}, l}}}
\newcommand{\vph}{\ensuremath{v_{\rm ph}}}
\newcommand{\vgr}{\ensuremath{v_{\rm gr}}}
\newcommand{\rhoi}{\ensuremath{\rho_{\rm i}}}
\newcommand{\rhoe}{\ensuremath{\rho_{\rm e}}}
\newcommand{\vai}{\ensuremath{v_{\rm Ai}}}
\newcommand{\vae}{\ensuremath{v_{\rm Ae}}}
\newcommand{\vgmin}{\ensuremath{v_{\rm gr}^{\rm min}}}
\newcommand{\omgmin}{\ensuremath{\omega^{\rm min}}}

\newcommand{\mathd}{\ensuremath{{\rm d}}}

\usepackage{color}

\shorttitle{Impulsively Generated Sausage Waves in Nonuniform Tubes}
\shortauthors{Yu et al.}

\begin{document}


\title{IMPULSIVELY GENERATED WAVE TRAINS IN CORONAL STRUCTURES:
I. EFFECTS OF TRANSVERSE STRUCTURING ON SAUSAGE WAVES IN PRESSURELESS TUBES}

\author{Hui Yu\altaffilmark{1}}
\author{Bo Li\altaffilmark{1}}
    \email{bbl@sdu.edu.cn}
\author{Shao-Xia Chen\altaffilmark{1}}
\author{Ming Xiong\altaffilmark{2}}
\and
\author{Ming-Zhe Guo\altaffilmark{1}}

\altaffiltext{1}{Shandong Provincial Key Laboratory of Optical Astronomy and Solar-Terrestrial Environment,
     Institute of Space Sciences, Shandong University, Weihai 264209, China}
\altaffiltext{2}{State Key Laboratory of Space Weather, National Space Science Center,
Chinese Academy of Sciences, Beijing 100190, China}

\begin{abstract}
The behavior of the axial group speeds of trapped sausage modes plays { an important} role
    in determining impulsively generated wave trains,
    which have often been invoked to account for
    quasi-periodic signals with quasi-periods of order seconds
    in a considerable number of coronal structures.
We conduct a comprehensive eigenmode analysis, both analytically and numerically,
    on the dispersive properties of sausage modes in pressureless tubes
    with three families of continuous { radial} density profiles.
We find a rich variety of the dependence on the axial wavenumber $k$ of
    the axial group speed $v_{\rm gr}$.
Depending on the density contrast and profile steepness as well as
    on the detailed profile description, the $v_{\rm gr}-k$ curves either possess or do not possess
    cutoff wavenumbers, and they can behave in either a monotonical or non-monotonical manner.
With time-dependent simulations, we further show that
    this rich variety of the group speed characteristics heavily influences
    the temporal evolution and Morlet spectra of impulsively generated wave trains.
In particular, the Morlet spectra can look substantially different from ``crazy tadpoles''
    found for the much-studied discontinuous density profiles.
We conclude that it is necessary to re-examine available high-cadence data
    to look for the rich set of temporal and spectral features,
    which can be employed to discriminate between the unknown forms
    of the density distributions transverse to coronal structures.
\end{abstract}

\section{INTRODUCTION}
\label{sec_intro}
Waves and oscillations in the magnetohydrodynamic (MHD) regime have been abundantly observed
    in the highly structured solar atmosphere
    \citep[see e.g.,][for recent reviews]{2005LRSP....2....3N,2007SoPh..246....3B,2012RSPTA.370.3193D,2016GMS...216..395W}.
Apart from being important agents for atmospheric heating
    \citep[see e.g.,][for some recent reviews]{2006SoPh..234...41K, 2012RSPTA.370.3217P,  2015RSPTA.37340261A, 2015RSPTA.37340269D},
    these waves and oscillations can also be exploited in the context of solar magneto-seismology (SMS) to
    infer the atmospheric parameters that are difficult to directly measure
    (for early ideas that laid the foundation of this practice, see
    \citeauthor{1970PASJ...22..341U}~\citeyear{1970PASJ...22..341U},
    \citeauthor{1970A&A.....9..159R}~\citeyear{1970A&A.....9..159R},
    \citeauthor{1975IGAFS..37....3Z}~\citeyear{1975IGAFS..37....3Z},
    \citeauthor{1984ApJ...279..857R}~\citeyear{1984ApJ...279..857R};
    also see recent reviews by e.g.,
    \citeauthor{2005LRSP....2....3N}~\citeyear{2005LRSP....2....3N},
    \citeauthor{2008IAUS..247....3R}~\citeyear{2008IAUS..247....3R},
    \citeauthor{2011SSRv..158..167E}~\citeyear{2011SSRv..158..167E},
    \citeauthor{2016SSRv..200...75N}~\citeyear{2016SSRv..200...75N}).
For instance, ``high-frequency'' fast waves with periods
    of order seconds were theoretically shown to play an important role
    in heating the solar corona~\citep[e.g.,][]{1994ApJ...435..482P, 1994ApJ...435..502P}.
On the observational side, quasi-periodic signals with short periods
    in coronal emissions presumably from density-enhanced loops
    have been known since the 1960s
    (e.g., \citeauthor{1969ApJ...155L.117P}~\citeyear{1969ApJ...155L.117P},
    \citeauthor{1969ApJ...158L.159F}~\citeyear{1969ApJ...158L.159F},
    \citeauthor{1970A&A.....9..159R}~\citeyear{1970A&A.....9..159R},
    \citeauthor{1973SoPh...32..485M}~\citeyear{1973SoPh...32..485M};
    see Table 1 of
    \citeauthor{1999ApJ...520..880A}~\citeyear{1999ApJ...520..880A}
    for a comprehensive compilation of measurements prior to 2000).
A substantial fraction of these measurements were made in optical passbands at total eclipses,
    and the measured quasi-periods range from
       $0.5-4$ seconds \citep[e.g.,][]{1984SoPh...90..325P, 1987SoPh..109..365P}
    to $4-7$ seconds \citep{2001MNRAS.326..428W,2002MNRAS.336..747W,2003A&A...406..709K}
    to $6-25$ seconds ~\citep{2016SoPh..291..155S}.
In other passbands, quasi-periodic pulsations (QPPs)  with similar periods
    have been found in the lightcurves of solar flares as measured with such instruments as
    the Nobeyama Radioheliograph \citep[NoRH, e.g.,][]{2001ApJ...562L.103A,  2003A&A...412L...7N, 2005A&A...439..727M, 2013SoPh..284..559K},
    SDO/AIA \citep[e.g.,][]{2012ApJ...755..113S},
    and more recently with the Interface Region Imaging Spectrograph \citep[IRIS,][]{2016ApJ...823L..16T}.
Much of the past and ongoing efforts in observing and understanding QPPs
    is summarized in the recent reviews by \citet{2009SSRv..149..119N} and \citet{2016SoPh..291.3143V}.

One needs to first identify the generation mechanisms of the above-mentioned
    short-period signals before they can be seismologically exploited.
As proposed by~\citet{1983Natur.305..688R,1984ApJ...279..857R},
    one mechanism is connected to fast sausage modes
    trapped in coronal tubes.
The key here is wave dispersion.
Consider the simplest case where the plasma beta is smaller than unity everywhere
    and coronal tubes are modeled by straight, field-aligned cylinders with { radial} density distributions
    in a top-hat fashion
    \citep[e.g.,][]{1978SoPh...58..165M, 1982SoPh...75....3S,1983SoPh...88..179E,1986SoPh..103..277C,2007AstL...33..706K, 2014ApJ...781...92V}.
Let $\vai$ and $\vae$ denote the internal and external Alfv\'en speeds, respectively
    ($\vae > \vai$).
Fast trapped sausage modes correspond to the axisymmetric perturbations whose
    axial phase speeds $\vph$ decrease with $k$ from $\vae$ to $\vai$.
There are an infinite number of trapped modes, characterized by
    the { radial harmonic number} $l$ ($l=1, 2, 3, \cdots$) such that the spatial distribution
    of the corresponding eigen-functions become increasingly complex with $l$.
{ (By convention, $l=1$ refers to the fundamental radial mode, while $l \ge 2$ refers to
    its radial harmonics.)
}    
For each $l$, trapped solutions exist only when the axial wavenumber $k$
    exceeds a critical value $\kcl$, which increases with $l$.
Focusing on the { case where} $l=1$, the wavenumber dependence
    of the axial group speed $\vgr$ is such that when $k$ increases from $k_{\rm c, 1}$,
    $\vgr$ first decreases to a local minimum $\vgr^{\rm min}$ before increasing towards $\vai$.
Given that the angular frequency $\omega$ increases monotonically with $k$,
    a similar non-monotonical behavior is also true for the $\vgr-\omega$ curves.
\citet{1984ApJ...279..857R} predicted that in response to an impulsive internal axisymmetric source,
    the signal in the tube at a distance $h$ sufficiently far away
    comprises three phases
    \citep[see also][for a heuristic discussion]{1986NASCP2449..347E}.
When $h/v_{\rm Ae} < t < h/v_{\rm Ai}$, individual wavepackets with progressively low group speeds arrive consecutively.
When $h/v_{\rm Ai} < t < h/\vgmin$, two wavepackets with the same group speed but different frequencies
    arrive simultaneously.
The signal now is expected to be stronger than in the first phase due to the superposition of multiple wavepackets.
When $t> h/\vgmin$, no incoming wavepackets are expected, resulting in a decaying signal
    oscillating at an angular frequency $\omega^{\rm min}$ where $\vgmin$ is attained.
These three phases are traditionally termed the periodic, quasi-periodic and decay (or Airy) phases.
The periods in the quasi-periodic phase are close to $2 \pi/\omgmin$,
    and therefore on the order of seconds because $\omgmin \sim \vai/R$ where $R$ is the tube radius.

The expected behavior of impulsively generated sausage waves
    has been shown to be robust
    when the density distribution transverse to coronal tubes
    is either discontinuous \citep{1996ApJ...472..398B, 2015ApJ...806...56O}
    or continuous \citep{2004A&A...422.1067S, 2015ApJ...814..135S}.
In addition, it takes place for the slab geometry as well
    \citep{1993SoPh..144..101M, 1994SoPh..151..305M, 2004MNRAS.349..705N, 2012A&A...537A..46J, 2012A&A...546A..49J, 2013A&A...560A..97P, 2016ApJ...826...78Y}.
What these computations suggest is that while
   such factors as the axial extent of the initial perturbation
   \citep{2015ApJ...806...56O} and density profile steepness \citep{2004MNRAS.349..705N,2015ApJ...814..135S}
   can be important, the temporal behavior of wave trains
   can still to a large extent be understood in the framework proposed by
   \citet{1984ApJ...279..857R} and \citet{1986NASCP2449..347E}.
In particular, the numerical studies by \citet{2004MNRAS.349..705N} and \citet{2015ApJ...814..135S}
   showed that the period and amplitude modulations in the wave trains transform into
   Morlet spectra characterized by a narrow tail followed by a broad head.
Given their shape, these spectra are called ``crazy tadpoles'',
   and were indeed found in both radio \citep[e.g.,][]{2010ITPS...38.2243J, 2013A&A...550A...1K}
   and optical measurements \citep[e.g.,][]{2003A&A...406..709K, 2016SoPh..291..155S}.
If the measured short-period signals can be interpreted this way,
   then their temporal and wavelet features can be exploited
   to yield such key information as the internal Alfv\'en speed, density contrast between loops and their surroundings,
   as well as the location of the impulsive source
   \citep[e.g.,][]{1984ApJ...279..857R, 2008IAUS..247....3R}.

Two features of typical $\vgr-\omega$ curves are { important} for the Morlet spectra to
   look like ``crazy tadpoles'' { (see also the discussions in Section~\ref{sec_conc})}.
One is the existence of cutoff wavenumbers,
   and the other is that the group speed curves possess a local minimum.
Restrict ourselves, for now, to the much-studied cases where primarily the { fundamental radial} modes are excited by an impulsive driver
   that is not too localized.
The narrow tail can then be explained by that the frequency varies little when $\vgr$ decreases from the external Alfv\'en speed
   to the local minimum.
Likewise, the broad head can be interpreted as the enhancement of Morlet power due to the superposition of multiple wavepackets
   with group speeds in the neighborhood of this minimum.
However, these two features largely pertain to the idealized cases where the transverse density distribution is discontinuous.
While the particular continuous profiles adopted by \citet{2004A&A...422.1067S} and \citet{2015ApJ...814..135S} yield wave trains largely compatible
   with the prediction by~\citet{1984ApJ...279..857R},
   this may not always be true if other profiles are employed.
A recent analytical study by \citet[][hereafter LN15]{2015ApJ...810...87L} showed that
   cutoff wavenumbers exist only when the { radial} density profiles drop sufficiently rapidly with distance.
Furthermore, our recent study \citep[][hereafter paper I]{2016ApJ...833...51Y} showed that the detailed form of the { radial} density profile
   close to the tube axis can result in a rich variety of $\vgr-\omega$ curves.
In addition to the classical case where a single local minimum exists, the curves
   can posses either more than one extrema or no extremum at all.
One then expects that overall four categories of the group speed curves are expected, based on whether
   cutoff wavenumbers exist and whether the $\vgr-\omega$ (or equivalently $\vgr-k$) curves
   behave in a non-monotonical fashion.
The present study aims to examine the observational consequences of the different behavior of the $\vgr-k$ curves.
To simplify our treatment, we will work in the cold MHD framework and
   examine three families of { radial} density profiles so chosen that
   all categories of the group speed curves result.
All profiles are characterized by only two parameters, one being a steepness parameter
   and the other being the density contrast.
For each family of profiles, we will first present an eigenmode analysis to construct group speed curves,
   which will then aid the interpretation of our numerical results from time-dependent computations.

This manuscript is organized as follows.
Section~\ref{sec_method_sol} starts with a description of the equilibrium configuration,
   and then presents the governing equations together with their solution methods.
Our methodology for studying impulsively generated wave trains is illustrated
   in Section~\ref{sec_numres_tophat} where the much-studied top-hat profiles are examined.
The purpose is to offer a reference case with which the results for other profiles can be compared.
Sections~\ref{sec_numres_monomu} to \ref{sec_numres_innermu} then examine the three families of density profiles in substantial detail.
Both the group speed curves obtained from an eigenmode analysis and
   the temporal behavior of wave trains found with time-dependent computations will be presented.
When mathematically tractable, an analytical approach will also be adopted to help understand the behavior
   of the group speed curves.
Finally, Sect.~\ref{sec_conc} closes this manuscript with our summary
    and some concluding remarks.

\section{GENERAL DESCRIPTION OF METHODS FOR EXAMINING IMPULSIVELY GENERATED SAUSAGE WAVES}
\label{sec_method_sol}

\subsection{Description of the Equilibrium Tube}
\begin{figure}
\centering
\includegraphics[width=0.7\columnwidth]{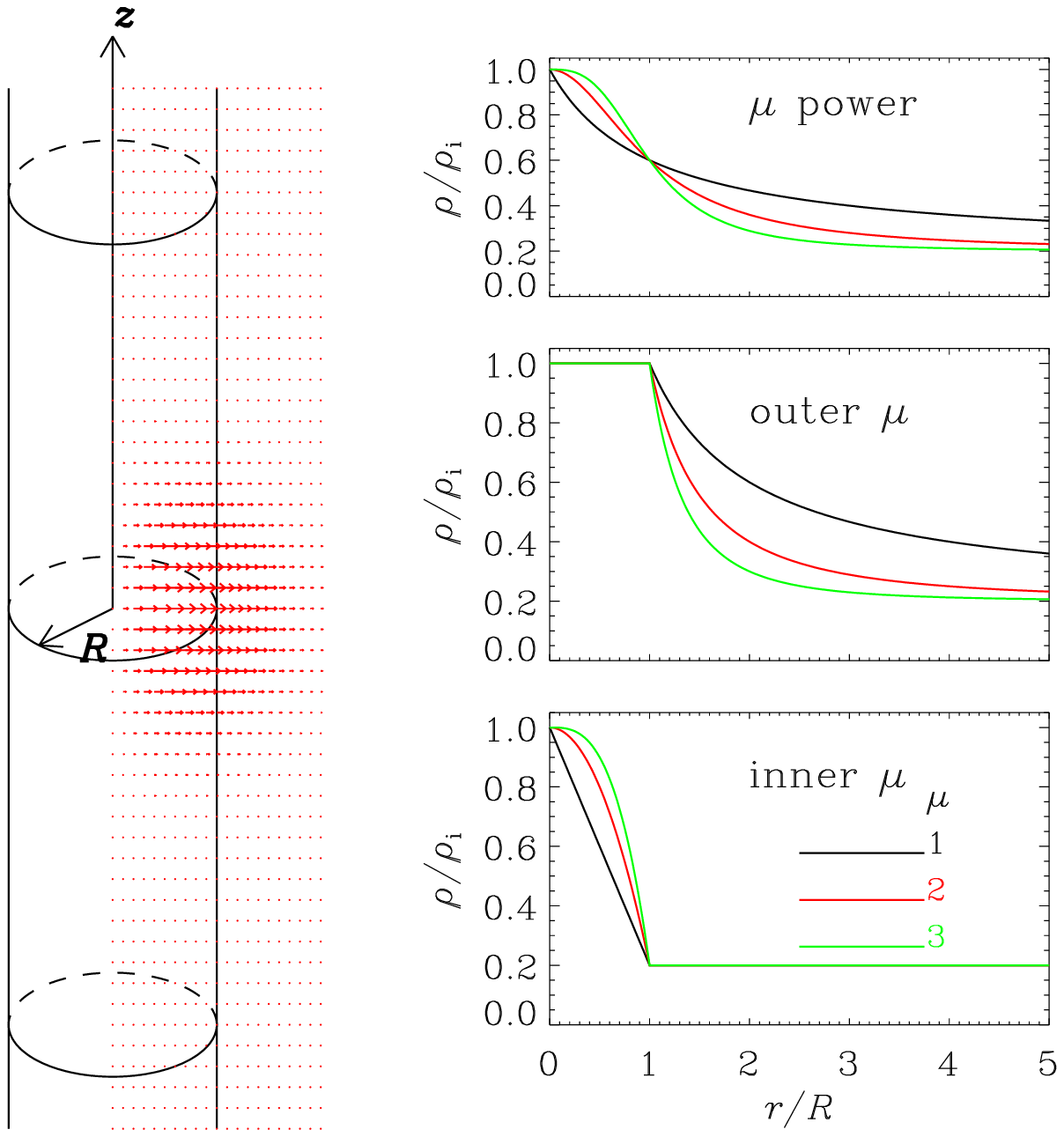}
\caption{
 Description of the modeled coronal tubes.
 In the left column, in addition to an illustration of the tubes,
     the initial perturbation to
     the radial velocity is also shown by the red arrows (see Equation~\ref{eq_IC_vr}).
 Shown in the right column are the three families of {radial} density profiles
     examined in this study.
 For illustration purposes, the density contrast $\rho_{\rm i}/\rho_{\rm e}$ is chosen to be $5$,
     while a number of different steepness parameters ($\mu$) are chosen as labeled.
}
 \label{fig_illus_profile}
\end{figure}

Appropriate for the solar corona, we work in the framework of cold MHD,
   for which the gas pressure is neglected and hence slow modes become irrelevant.
We consider sausage waves in a structured corona modeled by a density-enhanced straight tube with
   mean radius $R$ embedded in a uniform magnetic field ${\bf B}$.
Both ${\bf B}$ and the tube are directed in the $z$-direction
   in a cylindrical coordinate system $(r, \theta, z)$.
The left column of Figure~\ref{fig_illus_profile} offers an illustration
   of this setup.
The equilibrium density
   is assumed to be a function of $r$ only and of the form
\begin{equation}
 {\rho}(r)= \rhoe+(\rhoi-\rhoe) f(r),
    \label{eq_rho_profile_general}
\end{equation}
   where the function $f(r)$ decreases from unity at $r=0$ to zero when $r$ approaches infinity.
In other words, the equilibrium density decreases from $\rhoi$ at tube axis to $\rhoe$ far from
   the tube.
The Alfv\'en speed is given by $v_{\rm A} (r) = B/\sqrt{4\pi \rho (r)}$, which increases from $\vai$
   to $\vae$ with $v_{\rm A i, e} = B/\sqrt{4\pi \rho_{\rm i, e}}$.

While our methodology applies for arbitrary choices of $f(r)$, we nonetheless focus only on the following
   three families of profiles.
The first one is { called ``$\mu$ power'' and} given by
\begin{equation}
   f(r)= \displaystyle\frac{1}{1+\left(r/R\right)^{\mu}}~.
\label{eq_profile_monomu}
\end{equation}
The next one is called ``outer $\mu$'' and given by
\begin{equation}
   {f}(r)=\left\{
   \begin{array}{ll}
      1,     						& 0 \le r \le R, 			\\
      \left(r/R\right)^{-\mu}, 	& r \ge R.
   \end{array} \right.
\label{eq_profile_outermu}
\end{equation}
The third one is called ``inner $\mu$'' and in the form
\begin{equation}
   {f}(r)=\left\{
   \begin{array}{ll}
      1-\left(\displaystyle\frac{r}{R}\right)^\mu,     & 0 \le r \le R, \\
      0, 					& r \ge R.
   \end{array} \right.
\label{eq_profile_innermu}
\end{equation}
These profiles are illustrated in the right column of Fig.~\ref{fig_illus_profile} where
    $\rhoi/\rhoe$ is arbitrarily chosen to be $5$.
Evidently, $\mu$ is a measure of the profile steepness, and all profiles converge to a top-hat one
    when $\mu \rightarrow \infty$.
{ We will focus on coronal tubes with parameters typical of EUV active region (AR) loops,
   namely with density contrasts between $2$ and $10$ \citep[e.g.,][]{2004ApJ...600..458A}.
In addition, we will consider only $\mu \ge 1$ because otherwise the ``$\mu$ power'' and ``inner $\mu$'' profiles
   become cusped at $r=0$.
}    

\subsection{Governing Equations and Methods of Solution}

Now let $\delta \rho, \delta \vec{v}$ and $\delta \vec{b}$ represent the perturbations to the density,
    velocity, and magnetic field, respectively.
Working in the ideal cold MHD framework, and
    specializing to the axisymmetric case ($\partial/\partial\theta \equiv 0$),
    one finds that
    $\delta b_\theta, \delta v_\theta$, and $\delta v_z$ vanish.
Eliminating $\delta b_r$ and $\delta b_z$, one
    derives a single equation
    for $\delta v_r(r, z; t)$ \citep[e.g.,][]{2012ApJ...761..134N,2015SoPh..290.2231C,2015ApJ...812...22C}
\begin{equation}
\label{eq_vr_2ndorder}
 \frac{\partial^2 \delta v_r}{\partial t^2}
    =v_{\rm A}^2(r)
    \left(\displaystyle\frac{\partial^2}{\partial z^2}
         +\displaystyle\frac{\partial^2}{\partial r^2}
         +\displaystyle\frac{1}{r}\displaystyle\frac{\partial}{\partial r}
         -\displaystyle\frac{1}{r^2}\right)\delta v_r~.
\end{equation}
In addition, the density perturbation $\delta \rho$ is governed by
\begin{equation}
 \frac{\partial \delta \rho}{\partial t}
    =-\left[\delta v_r\frac{\partial \rho}{\partial r}
    +\frac{\rho}{r}\frac{\partial (r \delta v_r)}{\partial r}\right]~.
 \label{eq_rho}
\end{equation}

For each profile with a given combination of $[\rhoi/\rhoe, \mu]$, we will always start
   with an eigenmode analysis to establish the dispersive behavior of the trapped modes,
   the behavior of the group speed curves in particular.
Fourier-decomposing any perturbation $\delta g(r,z;t)$ as
\begin{eqnarray}
\label{eq_Fourier_ansatz}
  \delta g(r,z;t)={\rm Re}\left\{\tilde{g}(r)\exp\left[-i\left(\omega t-kz\right)\right]\right\}~,
\end{eqnarray}
   one finds from Eq.~(\ref{eq_vr_2ndorder}) that
\begin{eqnarray}
\label{eq_Fourier_xi}
    \frac{1}{r}\frac{\mathd}{\mathd r}\left(r\frac{\mathd \tilde{\xi}}{\mathd r}\right)
  +\left(\frac{\omega^2}{v^2_{\rm A}}-k^2-\displaystyle\frac{1}{r^2}\right)\tilde{\xi}=0~,
\end{eqnarray}
    where $\tilde{\xi} = i\tilde{v}_r/\omega$ is the Fourier
    amplitude of the radial Lagrangian displacement.
Equation~(\ref{eq_Fourier_xi}) can be formulated into a standard eigenvalue problem (EVP)
     when supplemented with the following boundary conditions (BCs)
\begin{eqnarray}
  \tilde{\xi} (r = 0) = 0~, \hspace{0.2cm}
  \tilde{\xi} (r \to \infty) \to 0~.
\label{eq_BVP_BC}
\end{eqnarray}
To solve this EVP,
    we employ a MATLAB boundary-value-problem solver BVPSuite in its eigen-value mode~\citep[see][for a detailed description of the code]{2009AIPC.1168...39K}.
In the appendix to \citet{2014A&A...568A..31L}, we have performed an extensive validation study of this code
    using available analytical solutions in the context of coronal seismology.
The end result is that the dimensionless angular frequency $\omega R/\vai$ can be formally expressed as
\begin{eqnarray}
    \frac{\omega R}{v_{\rm Ai}} = {\cal G}\left[kR; \rhoi/\rhoe, f(r)\right] .
\label{eq_omega_formal}
\end{eqnarray}
Note that both $\omega$ and the axial wavenumber $k$ are real-valued.
The axial phase and group speeds simply follow from the definitions
    $\vph = \omega/k$ and $\vgr = \mathd \omega/\mathd k$, respectively.

With the group speed curves at hand, we will then examine
    impulsively generated sausage waves by
    asking how a coronal tube responds to a localized initial perturbation.
For this purpose, we choose to solve Eq.~(\ref{eq_vr_2ndorder}), which is equivalent to, but simpler than,
    the linearized ideal cold MHD equations.
However, to initiate our simulations,
    an initial condition (IC) is needed for $\partial \delta v_r/\partial t$
    in addition to the IC for $\delta v_r$.
This can be found by noting that if we initially perturb only $\delta v_r$, then
\begin{eqnarray}
    \frac{\partial \delta v_r}{\partial t}(r, z; t=0) = 0~,
	    \label{eq_IC_vr_deriv}
\end{eqnarray}
    given the absence of $\delta b_r$ and $\delta b_z$ at $t=0$.
On the other hand, the following form is adopted as the initial radial velocity perturbation,
\begin{eqnarray}
   \delta v_r(r, z; t=0) = A{\rm e}^{1/2} v_{\rm Ai} \left(\frac{r}{\sigma_r}\right)
     \exp\left(-\frac{r^2}{2 \sigma_r^2}\right)
     \exp\left(-\frac{z^2}{2 \sigma_z^2}\right) ,
	    \label{eq_IC_vr}
\end{eqnarray}
    which ensures the parity of the generated wave trains by not displacing the tube axis.
A constant $\exp(1/2)$ is introduced such that initially $\delta v_r$ in units of $\vai$ attains a maximum of
   $A$, which is arbitrarily chosen to be unity.
Furthermore, $\sigma_r$ and $\sigma_z$ determine the extent to which the initial perturbation spans in
    the { radial} and axial directions, respectively.
Throughout this study, we choose $\sigma_r = \sigma_z = R/\sqrt{2}$ to ensure that the initial perturbation
    is neither too localized nor too extended.
This perturbation is shown by the red arrows in the left column of Fig.~\ref{fig_illus_profile}.

Having specified the ICs, we then develop a simple finite-difference (FD) scheme
    to numerically solve Eq.~(\ref{eq_vr_2ndorder})
    in a computational domain extending from $0$ to $L_r$ ($-L_z/2$ to $L_z/2$)
    in the $r$-($z$-) direction.
A uniform grid spacing $\Delta z = 0.08 R$
    is adopted in the $z$-direction for simplicity.
However, to speed up our computations, the grid points in the $r$-direction are chosen to be
    nonuniformly distributed.
We fix the spacing at $\Delta r_1 = 0.02 R$ when $r \le 3 R$.
We then allow it to increase in the way
    $\Delta r_{j+1} = 1.025 \Delta r_{j}$ with $j$ being the grid index
    until the spacing reaches $\sqrt{\rho_{\rm i}/\rho_{\rm e}} \Delta r_1$,
    beyond which it remains a constant again.
Despite this complication, we make sure that the FD approximations to
    the spatial derivatives in Eq.~(\ref{eq_vr_2ndorder}) are second order accurate.
Time integration is implemented with the leap-frog method, for which a uniform time
    step $\Delta t = 0.6 \Delta_{\rm min}/v_{\rm A, max}$ is adopted to comply with
    the Courant condition to ensure numerical stability.
Here $\Delta_{\rm min}$ ($v_{\rm A, max}$) represents the smallest (largest) value that
    the grid spacing (Alfv\'en speed) attains in the computational domain.
At the left boundary $r = 0$, we fix $\delta v_r$ at zero in view of the symmetric properties
    of sausage modes.
We choose both $L_r$ and $L_z$ to be sufficiently large such that the perturbations will not have sufficient time
    to reach the rest of the boundaries (at $r = L_r$ and $z = \pm L_z/2$).
In this regard, the boundary conditions there are irrelevant.
Nonetheless, we fix $\delta v_r$ at $0$ for simplicity in practice.
{ Furthermore, $L_r$ and $L_z$ are chosen to be $250 R$ and $500 R$, respectively.}

Given that available measurements attributed to impulsively generated sausage waves are primarily made
    with imaging rather than spectroscopic instruments,
    density perturbations will be more relevant compared with velocity perturbations.
In fact, $\delta \rho$ can be readily evaluated with Eq.~(\ref{eq_rho}), which is also solved with a scheme second order accurate
    in both time and space.
To initiate this evaluation, we note that $\delta \rho(r, z; t=0)=0$ because no density perturbation is applied.
The computed density perturbations are then sampled
   at a point along the tube axis and some distance $h$ away from the impulsive source.
A value of $h = 75~R$ is chosen such that only trapped modes play a role
   in determining $\delta\rho (r=0, z=h, t)$.

To summarize at this point, for each family of density profiles, we will adopt the following procedure to examine
    the signatures of impulsively generated wave trains.
\begin{enumerate}
 \item
    The $\vgr-k$ curves will be found by solving the EVP
      (Eqs.~\ref{eq_Fourier_xi} and \ref{eq_BVP_BC}) with BVPSuite.
    We will also present analytical expressions
       for $\vgr$ in some limiting cases to help better understand
       the behavior of the group speed curves.
    In particular, we will experiment with different combinations of $[\rhoi/\rhoe, \mu]$
       to see whether cutoff wavenumbers exist and whether the $\vgr-k$ curves
       behave in a non-monotonical fashion.
 \item
    We will solve Eq.~(\ref{eq_vr_2ndorder}) with the initial conditions (\ref{eq_IC_vr_deriv}) and (\ref{eq_IC_vr}), for which
       we fix $[A, \sigma_r, \sigma_z]$ at $[1, R/\sqrt{2}, R/\sqrt{2}]$.
    Equation~(\ref{eq_rho}) is simultaneously solved at each time-step when Eq.~(\ref{eq_vr_2ndorder}) is advanced.
    A time series corresponding to the density perturbation sampled at a specified location,
       $\delta\rho (r=0, z=75~R, t)$,
       will be subsequently analyzed with wavelet analysis,
       which is placed in the context of the $\vgr-k$ curves.
\end{enumerate}

Before proceeding, we note that
      whether cutoff wavenumbers exist is solely determined simply by the
      asymptotic behavior of $f(r)$ when $r$ is large.
With the aid of Kneser's oscillation theorem, LN15 established a general result that
      cutoff wavenumbers are present (absent) if $f(r)$ is steeper (less steep) than $r^{-2}$ at large distances.
However, solving the EVP is necessary to find out whether $\vgr$
      possesses a non-monotonical dependence on $k$.

\section{CORONAL TUBES WITH TOP-HAT PROFILES AS A REFERENCE CASE}
\label{sec_numres_tophat}
Let us start with the simplest case where the density distribution transverse to coronal tubes
    is discontinuous.
This is a limiting case that happens when $\mu\rightarrow \infty$
    for all of the profiles given by Equations~(\ref{eq_profile_monomu})
    to (\ref{eq_profile_innermu}).
The behavior of group speed curves is actually well-known
    \citep{1983Natur.305..688R,1984ApJ...279..857R,1988A&A...192..343E}.
It is nonetheless presented here to illustrate our procedure for examining
    impulsively generated waves.
To our knowledge, while it has been a common practice to interpret the temporal behavior
    of wave trains in the framework proposed by \citet{1983Natur.305..688R},
    the wavelet spectra have not been directly placed in the context
    of the $\vgr-k$ curves   \citep[cf.,][]{2015ApJ...806...56O,2015ApJ...814..135S}.

\subsection{Analytical Expectations}

For top-hat profiles,
    the solution to Equation~(\ref{eq_Fourier_xi}) in the uniform interior (exterior) is
    proportional to $J_1(nr)$ ($K_1(mr)$), where
\begin{eqnarray}
  n^2 = \displaystyle\frac{\omega^2}{\vai^2} - k^2~, \hspace{0.5cm}
  m^2 = k^2 - \displaystyle\frac{\omega^2}{\vae^2}~,
\label{eq_def_nm}
\end{eqnarray}
    and $J_j$ and $K_j$ represent Bessel function of the first kind
    and modified Bessel function of the second kind, respectively (here $j=1$).
Both $n^2$ and $m^2$ are non-negative for trapped modes.
The dispersion relation (DR) reads \citep[e.g.,][]{1983SoPh...88..179E, 2007AstL...33..706K}
\begin{eqnarray}
    n \displaystyle\frac{J_0(n R)}{J_1(n R)}
 =- m \displaystyle\frac{K_0(m R)}{K_1(m R)} .
\label{eq_DR_tophat}
\end{eqnarray}

There are an infinite number of branches of solutions to
    Equation~(\ref{eq_DR_tophat}).
Ordered by the { radial harmonic number} $l$, all these branches possess a low wavenumber cutoff given by
\begin{equation}
   \kcl = \displaystyle\frac{j_{0, l}}{R\sqrt{\rhoi/\rhoe-1}}~,
\label{eq_kc_tophat}
\end{equation}
   where $j_{0, l}$ is the $l$-th zero of $J_0$ ($l=1, 2, \cdots$).
Evidently, $\kcl$ increases with $l$.
Furthermore, it is no surprise to see that a cutoff wavenumber exists for all {radial harmonic numbers},
    given that $f(r)$ is identically zero when $r>R$ and consequently steeper
    than $r^{-2}$.

It will also be informative to find out how $\vgr$ behaves at large $kR$.
{It turns out that \citep[e.g,][]{2014A&A...568A..31L, 2015ApJ...806...56O,2016RAA....16...92Y}}
\begin{eqnarray}
   \displaystyle\frac{\vgr^2}{\vai^2} \approx \frac{1}{1+{j_{1,l}^2}/{(kR)^2}}~,
\label{eq_vg_tophat_bigK}
\end{eqnarray}
   where $j_{1, l}$ is the $l$-th zero of $J_1$ ($l=1, 2, \cdots$).
This means that $\vgr$ should eventually approach $\vai$ from below when $kR$ increases.
Note that Equation~(\ref{eq_vg_tophat_bigK}) comes directly from the approximate behavior
   of $\vph$ for sufficiently large $kR$
\begin{equation}
   \displaystyle\frac{\vph^2}{\vai^2} \approx 1+\frac{j_{1,l}^2}{k^2 R^2}~.
\label{eq_vph_tophat_bigK}
\end{equation}

\subsection{Group Speed Curves}
\begin{figure}
\centering
\includegraphics[width=0.8\columnwidth]{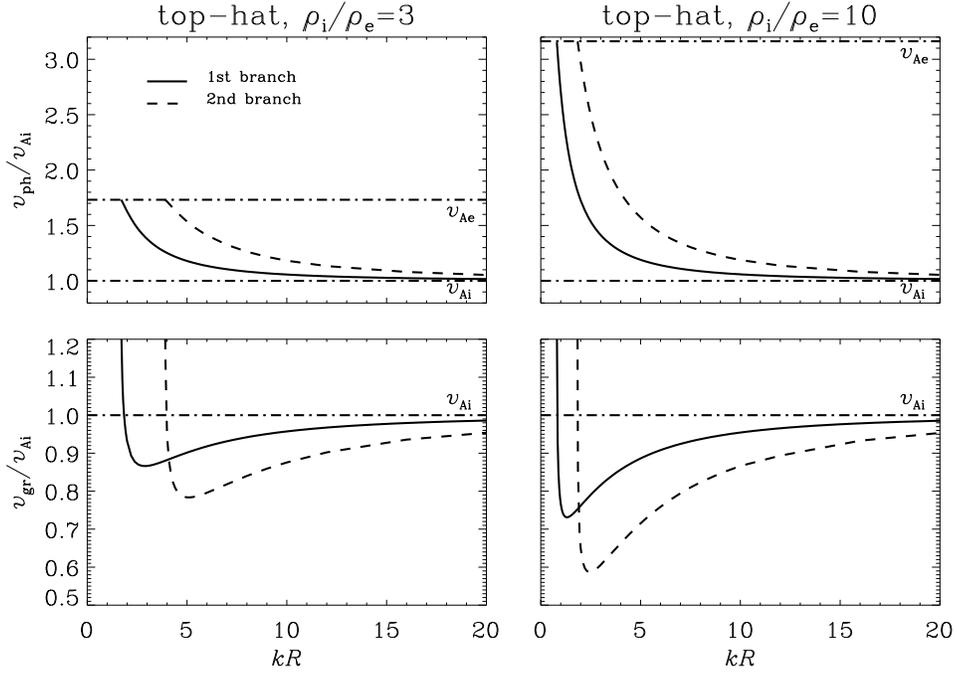}
 \caption{
 Dependence on the axial wavenumber $k$ of the axial phase (the upper row) and  group (lower) speeds
     for top-hat profiles with a density contrast of $3$ (the left column) and $10$ (right).
 The solid (dashed) curves represent the first (second) branch of trapped modes, corresponding to
     a {radial harmonic number} of $1$ ($2$).
 The horizontal dash-dotted lines represent the internal and external Alfv\'en speeds ($\vai$ and $\vae$).
}
 \label{fig_vphvg_k_tophat}
\end{figure}

Figure~\ref{fig_vphvg_k_tophat} presents the dependence on the axial wavenumber $k$
    of the axial phase (the upper row) and group (lower) speeds
    for both a density contrast of $3$ (the left column) and $10$ (right).
The solid and dashed curves correspond to { radial harmonic numbers} of $1$ and $2$, respectively.
These numerical results are found with BVPSuite and agree exactly with the numerical solutions
    to the analytical DR (\ref{eq_DR_tophat}).
One sees that cutoff wavenumbers exist for both density contrasts and for both branches,
    and $\kcl$ increases with $l$ but decreases with $\rhoi/\rhoe$
    as expected from Eq.~(\ref{eq_kc_tophat}).
Furthermore, all of the $\vgr-k$ curves show a non-monotonical dependence on $k$.
Letting $+$ ($-$) denote the tendency for $\vgr$ to increase (decrease) with $k$,
    this ``$-/+$'' behavior can be partly explained by the fact that
    asymptotically $\vgr$ approaches $\vai$ from below for arbitrary $l$
    and $\rhoi/\rhoe$.

\subsection{Temporal Evolution and Morlet Spectra of Density Perturbations}

Figure~\ref{fig_wavelet_tophat} displays the temporal evolution (the upper row)
    and the corresponding Morlet spectra (lower) of
    the density perturbations $\delta\rho$ sampled at a distance $h=75R$
    along the tube axis for both a density contrast $\rhoi/\rhoe = 3$ (the left column)
    and $10$ (right).
In the lower row, the left vertical axis represents the angular frequency $\omega$,
    and the right one represents the period $P$.
The Morlet spectra are created by using the standard wavelet toolkit devised by
    \citet{1998BAMS...79...61T}, and
    the dashed contours represent the $95\%$ confidence level
    computed by assuming a white-noise process for the mean background spectrum.
The dotted vertical lines correspond to the arrival times of wavepackets
    traveling at the internal ($\vai$) and external ($\vae$) Alfv\'en speeds.
Furthermore, the yellow curves represent $\omega$ as a function of $h/\vgr$ with the 
    { radial harmonic number}
    increasing from bottom to up.
These curves are essentially a replot of the curves in
    the lower row of Fig.~\ref{fig_vphvg_k_tophat}.

Consider the lower row first.
Placing the group speed curves on the Morlet spectrum indicates that
    the energy contained in the initial perturbation predominantly goes
    to the { fundamental radial} mode.
Furthermore, that the Morlet spectrum looks like  a ``crazy tadpole''
    is in close agreement with the reasoning by \citet{1986NASCP2449..347E}.
The narrow tail of the tadpole is threaded by the portion of the $\omega-h/\vgr$ curve
    where $\omega$ varies little.
On the other hand, the broad head occurs in the portion where $\vgr$ is close to its local minimum
    such that multiple wavepackets arrive simultaneously to yield
    some stronger Morlet power.
Comparing Figs.~\ref{fig_wavelet_tophat}b and \ref{fig_wavelet_tophat}d,
    one sees that this behavior is clearer for stronger density contrasts.
Now examining the upper row, one finds that
    wavepackets corresponding to higher { radial harmonics} also show up
    (see, e.g., the interval $t\gtrsim 110 R/\vai$ in Fig.~\ref{fig_wavelet_tophat}c).
They are simply too weak to discern in the Morlet spectra.

\section{CORONAL TUBES WITH ``$\mu$ power'' PROFILES}
\label{sec_numres_monomu}

The dispersive properties of sausage modes
    in coronal tubes with this profile were analytically examined in substantial detail by LN15,
    and some expectations can be found from the rather general analyses therein.
Let us first summarize what we expect
    regarding the cutoff wavenumbers and the dependence of the axial group speed on the axial wavenumber.

\subsection{Analytical Expectations}
LN15 showed that $\mu=2$ is a dividing line separating the cases where cutoff wavenumbers do not exist ($\mu <2$)
    from those where they do exist ($\mu >2$).
Care needs to be exercised when $\mu=2$, in which case LN15 showed that
    the cutoff wavenumber for { radial harmonic number} $l$ is given by
\begin{equation}
    \kcl = \displaystyle \frac{1}{R\sqrt{\rhoi/\rhoe-1}}~.
    \label{eq_monomu2_DR_cutoff}
\end{equation}
Note that $k_{{\rm c}, l}$ does not depend on $l$.
Our numerical computations with BVPSuite demonstrate that the expected behavior of cutoff wavenumbers indeed holds.

Some expectations on whether the $\vgr-k$ curves behave in a non-monotonical manner can
    be made by examining the asymptotic behavior of the group speeds at large axial wavenumbers.
LN15 showed that while in general the DR cannot be analytically established,
   it is possible to find an approximate one with the WKB approach when $\rhoi/\rhoe \gg 1$.
This approximate DR at large $kR$ yields that
\begin{equation}
  \displaystyle\frac{\vph^2}{\vai^2}
      \approx  1+\displaystyle\left(\frac{c}{kR}\right)^\beta ,
  \label{eq_monomu_vphbigK}
\end{equation}
   where $c$ is a constant that depends on $l$ and $\mu$, and $\beta = 2\mu/(\mu+2)$.
\footnote{
LN15 in their Eq.~(54) showed that the exponent $\beta$ reads $\mu/(\mu+2)$
    for $\mu <2$.
However, we believe this is a typo.
On the one hand, our numerical solutions with BVPSuite indicate that $\beta = 2\mu/(\mu+2)$
    for arbitrary values of $\mu$.
On the other hand, $\beta$ is expected to approach $2$ with increasing $\mu$
    because the profile becomes increasingly close to a top-hat one
    (see Eq.~\ref{eq_vph_tophat_bigK}).
}
It then follows that
\begin{eqnarray}
\displaystyle\frac{\vgr^2}{\vai^2}
    \approx
    1+\left(1-\beta\right)\left(\frac{c}{kR}\right)^\beta~.
\label{eq_monomu_vgrbigK}
\end{eqnarray}
Given that $\beta<1$ ($\beta >1$) when $\mu <2$ ($\mu >2$), one finds that
    $\vgr$ approaches $\vai$ from above (below) asymptotically.
Using the notations of ``$-$'' and ``$+$'', this means that asymptotically the $\vgr-k$ curves are of
    the ``$-$'' (``$+$'') type when $\mu <2$ ($\mu >2$).
The case with $\mu=2$ is special because while Eq.~(\ref{eq_monomu_vgrbigK}) suggests that
    $\vgr^2/\vai^2-1 \rightarrow 0$, it does not say whether $\vgr$ approaches $\vai$
    from above or below.
One way to tell this is to extend Eqs.~(\ref{eq_monomu_vphbigK}) and (\ref{eq_monomu_vgrbigK}) by incorporating terms
    of even higher order in $1/(kR)$.
Alternatively, we may employ full numerical solutions to the eigen-value problem by using BVPSuite.
These numerical solutions indicate that Eqs.~(\ref{eq_monomu_vphbigK}) and (\ref{eq_monomu_vgrbigK}) hold
    for arbitrary combinations of $[\rhoi/\rhoe, \mu]$.
In particular, they are true as long as $\rhoi/\rhoe >1$, way beyond the nominal range of applicability
    of the WKB analysis that assumes $\rhoi/\rhoe \gg 1$.
It is just that, in addition to $l$ and $\mu$,
    the constant $c$ depends on the density contrast as well.
When $\mu$ increases, this dependence gets weaker in that $c$ approaches $j_{1, l}$ as happens
    for top-hat profiles (see Eq.~\ref{eq_vph_tophat_bigK}).

\subsection{Group Speed Curves}

Figure~\ref{fig_vphvg_k_monomu} largely follows Fig.~\ref{fig_vphvg_k_tophat}
    in presenting the dependence on the axial wavenumber $k$
    of the axial phase and group speeds.
The difference is that now the parameter $\mu$ is relevant, and a number of different values
    for $\mu$ are examined as given by the curves in different colors.
Note that all of the curves are found with BVPSuite.
Consider cutoff wavenumbers first.
From both columns one sees that when $\mu < 2$ (the red and blue curves),
    both the first and second branches start at $k=0$.
In fact, our numerical solutions indicate that this holds for sausage modes
    of arbitrary { radial harmonic number} $l$.
On the contrary, when $\mu >2$ (the yellow and black curves), one sees
    that cutoff wavenumbers exist for sausage modes of any { radial harmonic number} $l$,
    and $\kcl$ increases with $l$.
In the particular case where $\mu=2$ (the green curves),
    $\kcl$ is non-zero but does not depend on $l$,
    as expected from Eq.~(\ref{eq_monomu2_DR_cutoff}).
Now examine the wavenumber dependence of the group speeds.
One sees that $\vgr$ behaves in a monotonical ``$-$'' (non-monotonical ``$-/+$'') manner
    when $\mu < 2$ ($\mu >2$), which can be partly explained by
    the asymptotic behavior of $\vgr$ as given by Eq.~(\ref{eq_monomu_vgrbigK}).
When $\mu=2$, the green curves indicate that $\vgr$ approaches $\vai$ from above at large wavenumbers.
In other words $\vgr^2/\vai^2-1 \rightarrow 0^+$,
    which cannot be directly deduced from Eq.~(\ref{eq_monomu_vgrbigK}).

{To sum up here,} the behavior of the group speed curves is determined solely by $\mu$.
Cutoff wavenumbers are absent (present) when $\mu < 2$ ($\mu \ge 2$),
    and the $\vgr -k$ curves behave in a ``$-$'' (``$-/+$'') manner when
    $\mu \le 2$ ($\mu > 2$).

\subsection{Temporal Evolution and Morlet Spectra of Density Perturbations}
Figure~\ref{fig_wavelet_monomu1} presents the temporal evolution and Morlet spectra for
    density perturbations sampled at a distance $h = 75 R$
    along the tube axis pertinent to a $\mu$ of $1$.
Note that the third branch corresponding to a {radial harmonic number} $l$ of $3$
    is also plotted in the lower row.
For the present choice of $\mu$, trapped modes exist for arbitrary wavenumbers
    and their group speeds {decrease} monotonically with $k$ from $\vae$ to $\vai$.
Consequently, two qualitative differences from the top-hat case arise.
First, the signals in the upper row do not extend beyond $h/\vai$.
Second, now the {radial harmonics} can be more readily excited because they can receive a substantial fraction
     of the energy contained in the initial perturbation.
This is particularly clear in Fig.~\ref{fig_wavelet_monomu1}d, where not only modes of {$l=2$}
     but those of $l=3$ can be clearly discerned.
Instead of tadpoles, the Morlet spectra look more like carps, for which the fins derive
     from modes of {radial harmonic number}s $l\ge 2$ and the bodies are obliquely directed
     due to the frequency drift of the modes of {$l=1$}.

Moving on to the cases with $\mu = 3$ as displayed in Fig.~\ref{fig_wavelet_monomu3},
     one sees that overall the Morlet spectra look similar to the top-hat cases.
However, in this case associating the strongest power with the superposition of multiple wavepackets
     is less clear, because the strongest power is not as closely related to the portion of
     the group speed curves embedding a local minimum (see e.g., Fig.~\ref{fig_wavelet_monomu3}b).
In addition, modes of {$l=2$} can now be more readily discerned: when $\rhoi/\rhoe=10$,
     these modes show up not only in the Morlet spectrum (Fig.~\ref{fig_wavelet_monomu3}d)
     but also in the temporal evolution
     (e.g., the interval between $20$ and $50 R/\vai$ in Fig.~\ref{fig_wavelet_monomu3}c).
Nonetheless, because cutoff wavenumbers exist and $\kcl$ increases with $l$, sausage modes of { $l=1$}
     remain dominant and account for the appearance of the narrow tails
     in the Morlet spectra.
The broad heads, however, turn out to be a result of the gradual frequency increase in the group speed curves beyond
     the speed minimum.

\section{CORONAL TUBES WITH ``OUTER $\mu$'' PROFILES}
\label{sec_numres_outermu}
This section examines coronal tubes with ``outer $\mu$'' profiles as given by Eq.~(\ref{eq_profile_outermu}).
For this family of profiles, closed-form DRs can be found for $\mu=1$ and $\mu=2$
   in much the same way that TE modes are examined in circular graded-profile optical fibers
   \citep[][p 270, Table 12-9]{1983optical..book....S}.
In what follows, we will start with an examination on these two particular choices of $\mu$
   and see what we can expect.

\subsection{Analytical Results for $\mu = 1$ and $\mu = 2$}
Let us start with the case where $\mu = 1$.
The solution to Eq.~(\ref{eq_Fourier_xi}) in the outer portion can be shown to have the form
\begin{eqnarray*}
     \tilde{\xi} \propto x^{-1/2} W_{\nu, 1}(x),
\end{eqnarray*}
    where $x= 2 m r$, and $W_{\nu, 1}(x)$ is Whittaker's W function \citep[][section 13.14]{NIST:DLMF}.
In addition,
\begin{eqnarray}
   \nu = \displaystyle\frac{kR}{2} \frac{\vph^2 (1-\rhoe/\rhoi)}{\vai^2\sqrt{1-\vph^2/\vae^2}} .
\end{eqnarray}
With $\tilde{\xi}$ expressible in terms of $J_1(n r)$ in the inner portion, the DR reads
\begin{eqnarray}
   nR \displaystyle\frac{J_0(n R)}{J_1(n R)}
 = mR - \nu + \displaystyle\frac{1}{2}-\displaystyle\frac{W_{\nu+1, 1}(2 mR)}{W_{\nu, 1}(2 mR)}~.
\label{eq_DR_outermu1}
\end{eqnarray}
We have used the fact that
\begin{equation}
 \displaystyle\frac{\mathd}{\mathd x} W_{\nu, 1}(x) =
   \left(\frac{1}{2}-\frac{\nu}{x}\right)W_{\nu, 1}(x)
   -\frac{W_{\nu+1, 1}(x)}{x}~.
\end{equation}

Some approximate results can be found in the limiting cases where $k\rightarrow 0$ or $k\rightarrow \infty$.
When $k\rightarrow 0$, it turns out that $\nu \rightarrow (2 l+1)/2$
   for { radial harmonic number} $l$ with $l = 1, 2, 3, \cdots$.
It then follows from the definition of $\nu$ that
\begin{equation}
   \displaystyle \vph
   \approx \vae \sqrt{ 1-\left(\frac{\rhoi/\rhoe-1}{2l+1}\right)^2 \left(kR\right)^2}~.
   \label{eq_outermu1_vph_smallK}
\end{equation}
Consequently,
\begin{equation}
   \displaystyle \vgr
   \approx \vae \left[1-\frac{3}{2}\left(\frac{\rhoi/\rhoe-1}{2l+1}\right)^2 \left(kR\right)^2\right]~.
   \label{eq_outermu1_vgr_smallK}
\end{equation}
While the non-existence of cutoff wavenumbers is readily expected since $f(r)$ is asymptotically less steep than $r^{-2}$,
    Equations~(\ref{eq_outermu1_vph_smallK}) and (\ref{eq_outermu1_vgr_smallK}) offer explicit expressions for
    what happens when $kR$ is small.
When $kR$ approaches infinity, the right hand side (RHS) of Equation~(\ref{eq_DR_outermu1}) tends to infinity.
As happens in the top-hat case,
    the group  and phase speeds for trapped modes of { radial harmonic number} $l$
    can still be approximated by Equations~(\ref{eq_vg_tophat_bigK})
    and (\ref{eq_vph_tophat_bigK}), respectively.

Now consider the case where $\mu = 2$.
The solution to Eq.~(\ref{eq_Fourier_xi}) in the outer portion can be shown to have the form
\begin{eqnarray*}
     \tilde{\xi} \propto K_{\nu}( m r),
\end{eqnarray*}
    where $K_\nu$ is modified Bessel's function of the second kind with
\begin{eqnarray}
   \nu = \displaystyle\sqrt{1-\frac{\vph^2}{\vai^2}\left(1-\frac{\rhoe}{\rhoi}\right)\left(kR\right)^2} .
\end{eqnarray}
Now that the cutoff wavenumbers are given by Equation~(\ref{eq_monomu2_DR_cutoff}),
   one sees that $\nu \to 0$ at the cutoff because $\vph = \vae = \vai \sqrt{\rhoi/\rhoe}$.
Moving away from this cutoff, $\vph k$ increases with $k$, meaning that $\nu$ is purely imaginary for
   trapped sausage modes \citep[see][section 10.45, for a discussion of $K_\nu$ of imaginary order]{NIST:DLMF}.
With $\tilde{\xi}$ expressible in terms of $J_1(n r)$ in the inner portion, the DR reads
\begin{eqnarray}
   (n R) \displaystyle\frac{J_0(n R)}{J_1(n R)}
 = 1 - \nu - (m R)\displaystyle\frac{K_{\nu-1}(m R)}{K_{\nu}(m R)}~.
\label{eq_DR_outermu2}
\end{eqnarray}
As is the case for $\mu = 1$, the RHS of Equation~(\ref{eq_DR_outermu2}) grows unbounded
   when $kR$ approaches infinity.
This means that once again $\vph$ and $\vgr$ can be approximated by Equations~(\ref{eq_vph_tophat_bigK})
    and (\ref{eq_vg_tophat_bigK}), respectively.

\subsection{Group Speed Curves}

Figure~\ref{fig_vphvg_k_outermu} presents, in a form identical to Fig.~\ref{fig_vphvg_k_monomu},
    the wavenumber dependence of the phase and group speeds as
    found with BVPSuite.
We have also numerically solved the analytical DRs pertinent to $\mu=1$ and $2$ (Eqs.~\ref{eq_DR_outermu1} and \ref{eq_DR_outermu2}),
    and found that these solutions agree exactly with the red and green curves.
Note that these analytical results yield an asymptotic behavior for
    $\vph$ and $\vgr$ to depend on $k$ in the ways described by Eqs.~(\ref{eq_vph_tophat_bigK})
    and (\ref{eq_vg_tophat_bigK}), respectively.
In fact, this asymptotic behavior holds for arbitrary combinations of $[\rhoi/\rhoe, \mu]$ as found from an extensive set
    of computations, and partly explains why the $\vgr-k$ curves
    are exclusively of the $-/+$ type for ``outer $\mu$'' profiles.
Figure~\ref{fig_vphvg_k_outermu} is a subset of our computations and illustrates that with increasing wavenumber,
    $\vgr$ always decreases first to some minimum before eventually approaching $\vai$ from below.

When it comes to cutoff wavenumbers, one sees exactly the same behavior as in the { ``$\mu$ power''} cases.
Cutoff wavenumbers do not exist for any { radial harmonic number} $l$ when $\mu < 2$,
    but do exit and increase with $l$ when $\mu >2$.
In the particular case where $\mu=2$, cutoff wavenumbers exist but do not depend on
    the { radial harmonic number}.
They are actually exactly given by Equation~(\ref{eq_monomu2_DR_cutoff}).
The behavior of cutoff wavenumbers is much expected, because it is determined solely by
    the asymptotic $r$-dependence of $f(r)$, which agrees exactly with the { ``$\mu$ power''} case:
    $1/[1+(r/R)^\mu] \rightarrow (r/R)^{-\mu}$ when $r\rightarrow \infty$.

{ To sum up here, cutoff wavenumbers are absent (present) when $\mu < 2$ ($\mu \ge 2$)
    as happens for the ``$\mu$ power'' cases.
However, we note that in this case the $\vgr-k$ curves are unanimously
    of the ``$-/+$'' type, which is different from the { ``$\mu$ power''} cases
    where $\mu=2$ also separates the ``$-$'' from the ``$-/+$'' behavior.
}

\subsection{Temporal Evolution and Morlet Spectra of Density Perturbations}

Figure~\ref{fig_wavelet_outermu1} presents the temporal evolution and Morlet spectra of
    the sampled density perturbations for ``outer $\mu$'' profiles with $\mu=1$.
One sees that the Morlet spectra bear similarities to those in both Figs.~\ref{fig_wavelet_monomu1}
    and \ref{fig_wavelet_monomu3}: they look like crazy tadpoles with fins.
This is understandable because the characteristics of the $\vgr-k$ curves in the present case
    is a mixture of those for $\mu=1$ and $\mu=3$ in the { ``$\mu$ power''} cases.
To be specific, no cutoff wavenumbers exist for trapped modes of any { radial harmonic number} $l$,
    the consequence being that modes of $l\ge 2$ can be excited to account for the fins.
On the other hand, the non-monotonical wavenumber dependence of the group speeds
    explains why the { main bodies of the wavelet spectra 
    are in the shape of ``crazy tadpoles''}.

Moving on to the cases with $\mu=3$ as given in Fig.~\ref{fig_wavelet_outermu3},
    one sees the classical crazy tadpoles in the Morlet spectra.
This behavior agrees closely with the top-hat case, and we need only to mention that
    the superposition of multiple wavepackets plays a key role in shaping
    the broad head of the tadpoles.

\section{CORONAL TUBES WITH ``INNER $\mu$'' PROFILES}
\label{sec_numres_innermu}

This section examines ``inner $\mu$'' profiles.
In this case, Eq.~(\ref{eq_Fourier_xi}) in the inner portion is exactly solvable
    in terms of compact closed-form expressions when $\mu=2$.
\footnote{\citet[][Eq.~8]{1986NASCP2449..347E} presented a compact analytical DR for arbitrary $\mu$.
Our numerical solutions with BVPSuite indicate that this DR holds only when $\mu$ is sufficiently large,
    a point already pointed out in the context of optical fibers by \citet{1980ITMTT..28.1113O}.
For large $\mu$, however, this DR is not far from the one pertinent to top-hat profiles
    (Eq.~\ref{eq_DR_tophat}).
}

\subsection{Analytical Results for $\mu = 2$}
An analytical treatment for this particular $\mu$ was initially given by \citet{1965PhFl....8..507P}
    who assumed the ambient to be a vacuum ($\rhoe = 0$).
Our paper I offered a generalization by presenting a DR valid for arbitrary finite $\rhoe$ but focused only
    on the { fundamental radial} modes.
We now further exploit that DR to examine trapped modes of arbitrary { radial harmonic number}
    $l$ ($l=1, 2, 3, \cdots$).

The following definitions are necessary,
\begin{eqnarray*}
   && p = \displaystyle\frac{\omega R}{\vai}
      \sqrt{1-\displaystyle\frac{\rhoe}{\rhoi}} ,  \\
   && \alpha = 1-\displaystyle\frac{\left(\omega R/\vai\right)^2-\left(k R\right)^2}{4 p} .
\end{eqnarray*}
The solution to Eq.~(\ref{eq_Fourier_xi}) in the inner portion can be shown to have the form
\begin{eqnarray*}
     \tilde{\xi} \propto x^{1/2} \exp(-x/2) M(\alpha, 2, x),
\end{eqnarray*}
    where $x= p (r/R)^2$, and $M(a, b, x)$
    is Kummer's M function \citep[][section 13.2]{NIST:DLMF}.
With $\tilde{\xi}$ expressible in terms of $K_1(m r)$ in the outer portion, the DR then reads
\begin{eqnarray}
   - (m R) \displaystyle\frac{K_0(m R)}{K_1(m R)}
 = 2 - p + \alpha p \displaystyle\frac{M(\alpha+1, 3, p)}{M(\alpha, 2, p)}~.
\label{eq_DR_innermu2}
\end{eqnarray}

Approximate expressions can be found for both the cutoff wavenumbers and the asymptotic behavior of
    the phase and group speeds at large wavenumbers.
In fact, both are related to the fact that the DR for any { radial harmonic number} $l$
    can be approximated by $\alpha = 1-l$ for nearly the entire range of wavenumbers.
Given that $\vph = \vae$ at the cutoff, one sees that
\begin{eqnarray*}
&& p = \displaystyle \kc R \sqrt{\frac{\rhoi}{\rhoe}-1} ,  \\
&& \alpha = 1-\frac{\kc^2 R^2({\rhoi}/{\rhoe}-1)}{4p} = 1-\frac{p}{4} .
\end{eqnarray*}
This means that the cutoff can be approximated by $p\approx 4 l$, or equivalently
\begin{equation}
  k_{{\rm c}, l} R \approx \displaystyle\frac{4 l}{\sqrt{\rhoi/\rhoe-1}} .
  \label{eq_innermu2_cutoff}
\end{equation}
It turns out this approximation is increasingly accurate with $l$,
   and is accurate to { within  $13.7\%$} for $l=1$.
\footnote{Given that $mR \rightarrow 0$ at the cutoff, one readily recognizes that
   the left hand side (LHS) of the DR (\ref{eq_DR_innermu2}) approaches zero.
  By further noting that $\alpha = 1-p/4$, one can see the RHS as a function of $p$ only.
  Numerically solving this DR yields that $p= 3.52, 7.45, 11.42, \cdots$ for $l=1, 2, 3, \cdots$.
  The analytical expression $p\approx 4l$ overestimates the exact values by 
     $13.7\%$, $7.35\%$, and $5.04\%$ for $l=1$, $2$, and $3$, respectively. 
}
On the other hand, $\alpha$ at large wavenumbers  is almost exactly $1-l$, yielding a simple equation quadratic in $\omega$.
Solving this equation for $\omega$, one finds that
\begin{eqnarray*}
 \displaystyle\frac{\omega R}{\vai} = 2l \sqrt{1-\frac{\rhoe}{\rhoi}} + \sqrt{4l^2\left(1-\frac{\rhoe}{\rhoi}\right)+k^2 R^2} .
\end{eqnarray*}
Consequently,
\begin{eqnarray}
  \displaystyle\frac{\vph}{\vai}
  \approx 1+\displaystyle\frac{2l\sqrt{1-\rhoe/\rhoi}}{k R}
                +\displaystyle\frac{2 l^2 (1-\rhoe/\rhoi)}{(k R)^2} ,
  \label{eq_innermu2_bigK_vph}
\end{eqnarray}
   and
\begin{eqnarray}
  \displaystyle\frac{\vgr}{\vai}
  \approx 1-\displaystyle\frac{2 l^2(1-\rhoe/\rhoi)}{(k R)^2} .
  \label{eq_innermu2_bigK_vg}
\end{eqnarray}
In other words, one expects to see that $\vgr$ approaches $\vai$
   from below.

\subsection{Group Speed Curves}
With BVPSuite we have computed the phase and group speed curves for an extensive set of combinations of $[\rhoi/\rhoe, \mu]$.
We find that cutoff wavenumbers ($k_{{\rm c}, l}$) exist for arbitrary { radial harmonic number} $l$
   and $k_{{\rm c}, l}$ increases with $l$.
The existence of cutoff wavenumbers is expected because $f(r)$ is identically zero for $r>R$
    and therefore steeper than $r^{-2}$.
However, it is interesting to note that $\vgr$ at large wavenumbers follows exactly the same $k$ dependence
    as given by Eq.~(\ref{eq_monomu_vgrbigK}) for { ``$\mu$ power''} profiles.
Not only the exponent $\beta$ but also the constant $c$ are found to be identical.
In particular, for the particular choice $\mu = 2$ (and hence $\beta =1$), 
    we find that $\vgr^2/\vai^2-1 \rightarrow 0$ as expected from Eq.~(\ref{eq_monomu_vgrbigK}).
It is just that in the present case $\vgr^2/\vai^2-1 \rightarrow 0^{-}$ (see Eq.~\ref{eq_innermu2_bigK_vg}),
    in contrast to the { ``$\mu$ power''} case where
    $\vgr^2/\vai^2-1 \rightarrow 0^{+}$.
Then why is the asymptotic group speed behavior identical to what happens for { ``$\mu$ power''} profiles,
    barring this minor difference happening for $\mu=2$?
To understand this, we note that at large density contrasts,
    LN15 showed that the dispersive properties of sausage modes are determined by the behavior of the turning points
    of the function $Q(r)$ as given by Eq.~(44) therein.
For large wavenumbers, the locations of both turning points correspond to an $r$ substantially smaller than $R$.
In this portion, however, the ``inner $\mu$'' profiles are not too different from the { ``$\mu$ power''} ones because
    $1/[1+(r/R)^\mu] \approx 1-(r/R)^\mu$.
While the analytical analysis in LN15 was conducted in the WKB framework by assuming $\rhoi/\rhoe \gg 1$, our numerical results
    indicate that actually the asymptotic behavior holds for arbitrary values of $\rhoi/\rhoe$ as long as $\rhoi/\rhoe >1$.

Figure~\ref{fig_vphvg_k_innermu} presents, in the same format as Fig.~\ref{fig_vphvg_k_monomu}, some examples of
    the phase (the upper row)
    and group (lower) speed curves taken from the extensive set of numerical results.
Regarding cutoff wavenumbers, one sees that
    they exist for all the examined $\mu$ values and for both density contrasts.
In addition, one finds for both density contrasts that $\mu=2$ is a dividing line that separates
    the cases where $\vgr$ eventually approaches $\vai$ from above
    from those where $\vgr$ approaches $\vai$ from below.
As to the particular case where $\mu = 2$, one sees that $\vgr$ tends to $\vai$ from below as expected
    from Eq.~(\ref{eq_innermu2_bigK_vg}).
However, while overall the $\vgr-k$ curves behave in a ``$-/+$'' manner when $\mu \ge 2$,
    some interesting complications arise when $\mu < 2$.
For a combination $[\rhoi/\rhoe, \mu]$ of $[3, 1]$ (the red curves in the left column),
    $\vgr$ is found to be a monotonically decreasing function of $k$.
However, when $[\rhoi/\rhoe, \mu]=[10, 1]$, the red curves in the right column indicate that
    $\vgr$ decreases with $k$ first and then increases before eventually decreasing toward $\vai$.
In other words, the $\vgr-k$ curves are of the ``$-/+/-$'' type, possessing not only a local minimum
    but also a local maximum.
If examining the cases where $\mu = 1.5$, the blue curves in both columns suggest that
    the $\vgr -k$ curves both belong to this ``$-/+/-$'' category.

Specializing to the density contrasts in the range $[2, 10]$
    and $\mu$ in the range $[1, 3]$,
    Figure~\ref{fig_DRsum_innermu} summarizes the behavior of the group speed curves
    for ``inner $\mu$'' profiles.
One sees that in this case, cutoff wavenumbers exist regardless of $\rhoi/\rhoe$ or $\mu$.
While the $\vgr-k$ curves are unanimously of the ``$-/+$'' type for $\mu \ge 2$,
    the area $\mu <2$ is further subdivided into two regions separated by the left dash-dotted curve.
For combinations of $[\rhoi/\rhoe, \mu]$ that lie below (above) this curve,
    the $\vgr-k$ curves behave in a ``$-$'' (``$-/+/-$'') manner.

\subsection{Temporal Evolution and Morlet Spectra of Density Perturbations}
Figure~\ref{fig_wavelet_innermu1} presents the temporal evolution and Morlet spectra of density perturbations
    for two ``inner $\mu$'' profiles, both with $\mu=1$.
Common to both columns is that the { fundamental radial} modes dominate the signals.
Nonetheless, some substantial difference exists in that
    the Morlet spectrum for $\rhoi/\rhoe = 3$ (the left column) looks like a submarine whereas
    it recovers the classical shape of crazy tadpoles when $\rhoi/\rhoe = 10$ (right).
The former appearance is understandable because the energy contained in the initial perturbation
    is primarily distributed to the starting portion of the $\vgr-k$ curves where $\omega$
    shows little variation.
The appearance of a Morlet spectrum in a tadpole shape in the right column requires some explanation, because
    for this combination of $[\rhoi/\rhoe, \mu]$, the $\vgr-k$ curves are actually
    of the ``$-/+/-$'' type, which is different from the top-hat case.
In fact, the broad head can still be attributed to the superposition of multiple wavepackets with group speeds close to
    the local minimum.
Nonetheless, different from the tadpoles in 
    Figs.~\ref{fig_wavelet_tophat}d and \ref{fig_wavelet_outermu3}d, 
    the strongest Morlet power in the present tadpole
    does not extend beyond $h/\vai$.
This occurs because the local minimum in the $\vgr-k$ curve is larger than the internal Alfv\'en speed.

Figure~\ref{fig_wavelet_innermu3} displays what happens when $\mu=3$.
One sees the classical crazy tadpoles in the Morlet spectra, which is especially true
    when the density contrast is chosen to be $10$.
In fact, these spectra are very similar to those in Fig.~\ref{fig_wavelet_tophat},
    which is readily understandable given the similarities of the group speed curves
    in both sets of figures.

\section{SUMMARY}
\label{sec_conc}
Quasi-periodic propagating disturbances with quasi-periods of order seconds
    have been seen in a substantial number of
    coronal structures.
When interpreted as impulsively generated sausage wave trains as proposed
    by \citet{1983Natur.305..688R} and \citet{1984ApJ...279..857R},
    these disturbances can help yield such key information as
    the magnetic field strength in the structures
    as well as the density structuring transverse to magnetized structures.
An important piece of observational evidence corroborating this interpretation is
    the appearance of the pertinent Morlet spectra in the shape of ``crazy tadpoles''.
This appearance derives from two key features of the dependence on the axial wavenumber ($k$)
    of the axial group speeds ($\vgr$) of sausage modes trapped in coronal tubes.
One is the existence of cutoff wavenumbers, only beyond which can sausage modes be trapped.
The other is the non-monotonical behavior of the $\vgr-k$ curves, namely $\vgr$ first decreases rapidly
    with $k$ to some minimum before increasing again.
These two key features are largely found in theoretical studies of coronal tubes with discontinuous
    density profiles.
The aim of this study is to examine the effects of continuous { radial} density profiles
    on the $\vgr-k$ curves and consequently on the temporal and wavelet signatures
    of impulsively generated wave trains.
To this end, we worked in the framework of cold MHD and examined three families
    of density profiles, all of which are characterized by the density contrast $\rhoi/\rhoe$
    and a steepness parameter $\mu$.
Through analytical and numerical solutions to the relevant eigenvalue problem,
    for each family of profiles we categorized the $\vgr-k$ curves, which provided
    the context for the interpretation of the temporal evolution and Morlet spectra
    of wave trains obtained with numerical solutions to the time-dependent MHD equations.

In agreement with \citet{2015ApJ...810...87L}, we find that
    cutoff wavenumbers exist as long as the {radial} density profiles
    fall more rapidly than $r^{-2}$, where $r$ is the distance from the tube axis.
However, the $k$ dependence of $\vgr$ is determined not only by the combination $[\rhoi/\rhoe, \mu]$
    but also by how the density profile is described.
Let ``$-$'' (``$+$'') denote the tendency for $\vgr$ to decrease (increase) with $k$.
Overall the $\vgr-k$ curves can be grouped into four categories.
Category I corresponds to the classical case where cutoff {wavenumbers} exist and
    the $\vgr-k$ curves behave in a non-monotonical manner.
Typically the Morlet spectra look like ``crazy tadpoles'', which result
    even when the $\vgr-k$ curves behave in a ``$-/+/-$'' rather than the ``$-/+$'' fashion
    (see Fig.~\ref{fig_wavelet_innermu1}d).
Category II corresponds to the case where cutoff wavenumbers do not exist and
    $\vgr$ decreases monotonically with $k$.
The Morlet spectra are now in a shape of ``carps with fins'', where the fins result because
    {radial harmonics} can be readily excited.
While the majority of the Morlet power is still associated with the {fundamental radial modes},
    the corresponding spectra do not posses a broad head.
Category III pertains to case where cutoff wavenumbers do not exist but
    the $\vgr-k$ curves are in a ``$-/+$'' fashion.
In this case the Morlet spectra look like ``crazy tadpoles with fins'' (see Fig.~\ref{fig_wavelet_outermu1}),
    a mixture of the characteristics associated with categories I and II.
Category IV results when cutoff wavenumbers exist but the $\vgr-k$ curves
    are of the ``$-$'' type.
The Morlet spectra are now nearly symmetric and
    show no significant frequency modulation (see Fig.~\ref{fig_wavelet_innermu1}b).

\begin{figure}
\centering
\includegraphics[height=80mm]{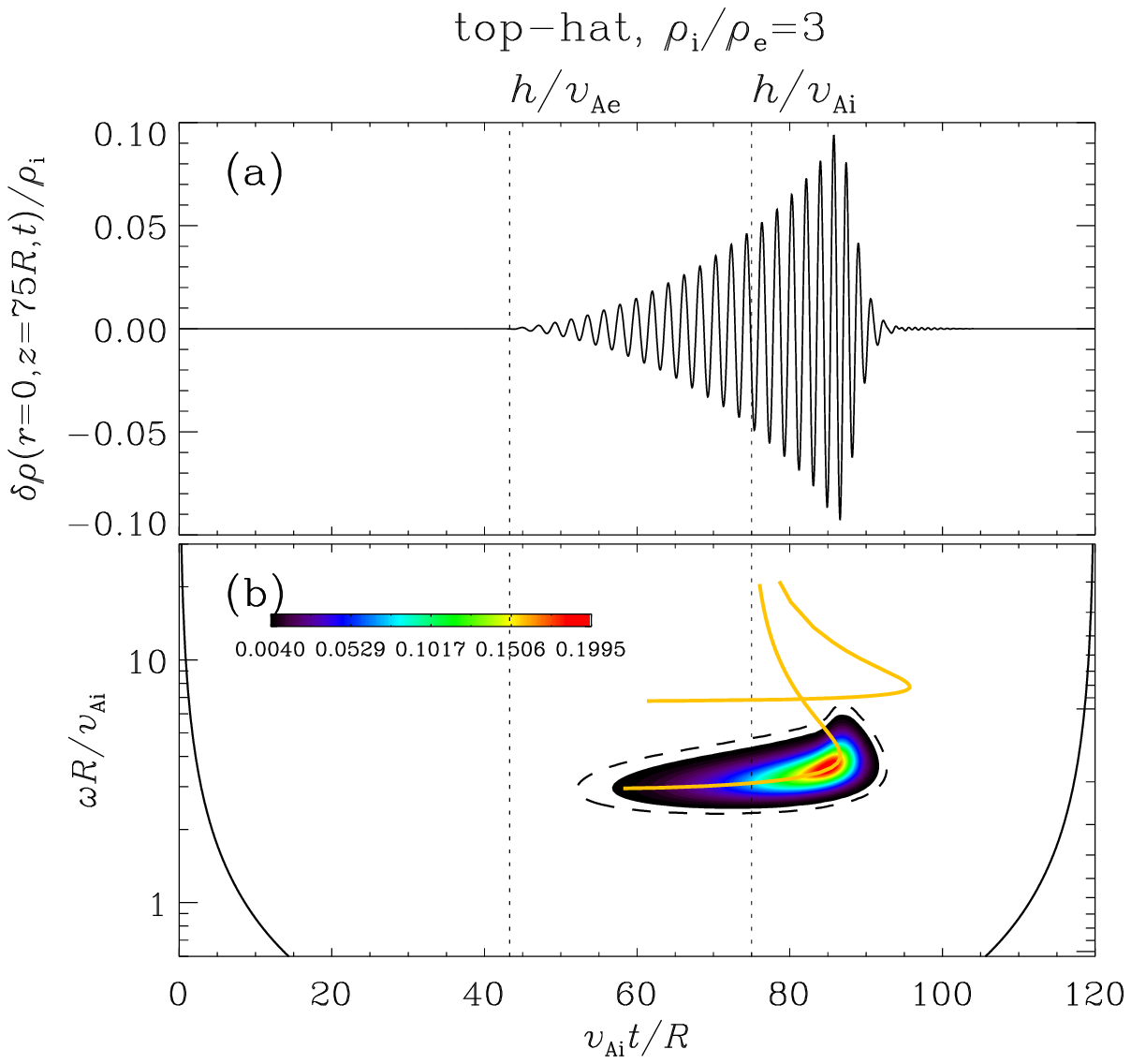}
\includegraphics[height=80mm]{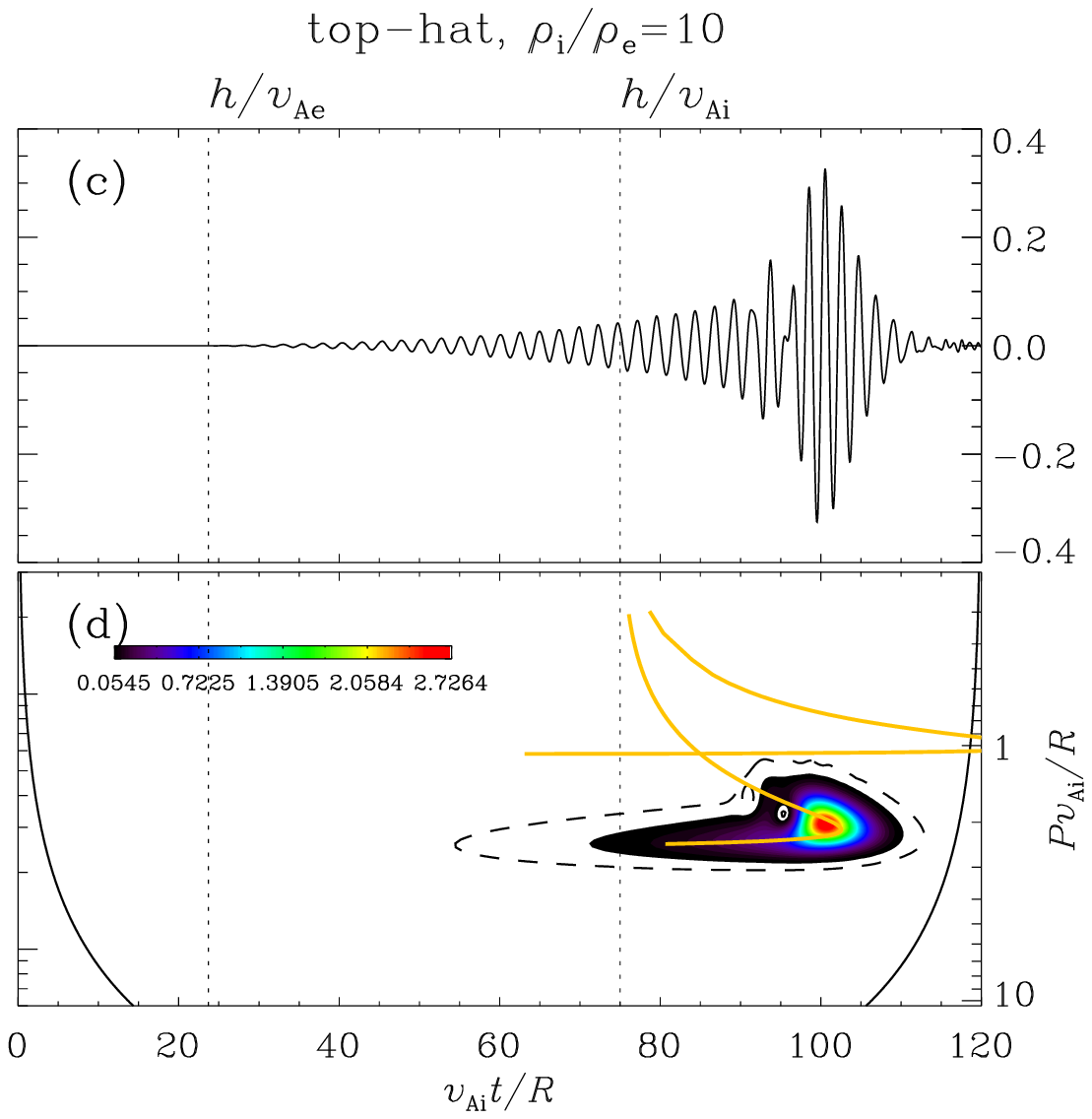}
\caption{
 Density perturbations $\delta\rho$ at a distance $h=75R$
     from the impulsive source along the axis of a coronal tube
     with top-hat profiles.
 The left and right columns pertain to a density contrast $\rho_{\rm i}/\rho_{\rm e}$
     of $3$ and $10$, respectively.
 In addition to the temporal evolution (the upper row),
     the corresponding Morlet spectra are also shown (lower).
 The left and right vertical axes in the lower row represent
     the angular frequency $\omega$ and period $P$, respectively.
 Furthermore, the black solid curves represent
     the cone of influence,
     and the area inside the dashed contour indicates where the Morlet power
     exceeds the $95\%$ confidence level.
 The dotted vertical lines correspond to the arrival times
     of wavepackets traveling at the internal and external Alfv\'en speeds as labeled.
 The yellow curves represent $\omega-h/\vgr$ as found from the eigenmode analysis,
     with the {radial harmonic number} increasing from bottom to top.
}
\label{fig_wavelet_tophat}
\end{figure}

\begin{figure}
\centering
\includegraphics[width=0.8\columnwidth]{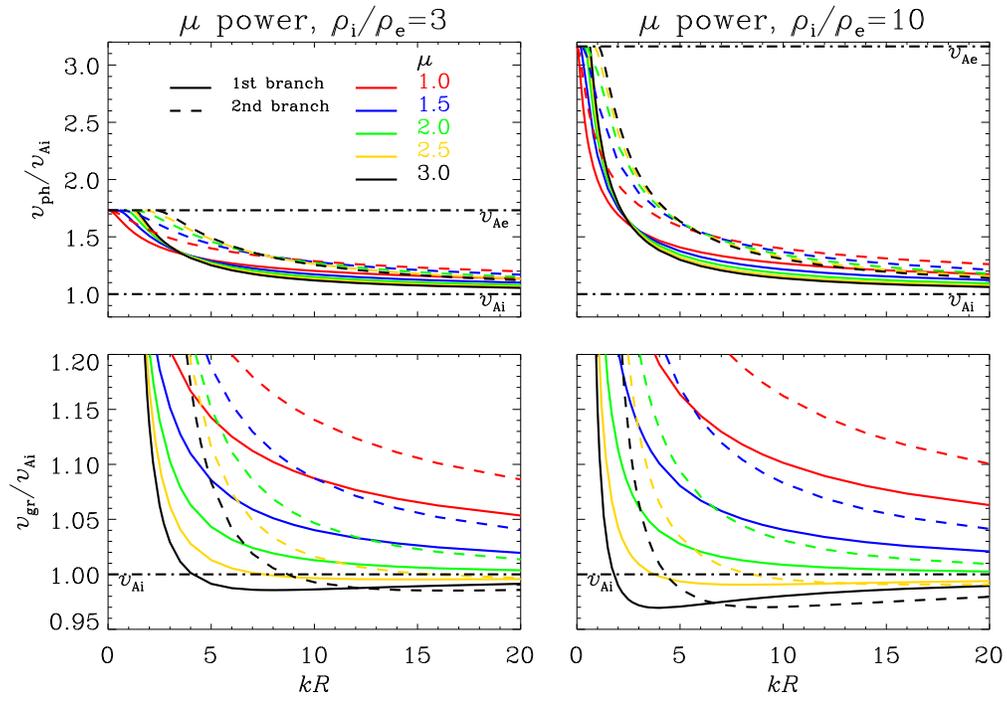}
 \caption{
Similar to Fig.~\ref{fig_vphvg_k_tophat} but for {``$\mu$ power''} profiles with a number of $\mu$ as labeled.
 }
 \label{fig_vphvg_k_monomu}
\end{figure}

\begin{figure}
\centering
\includegraphics[height=80mm]{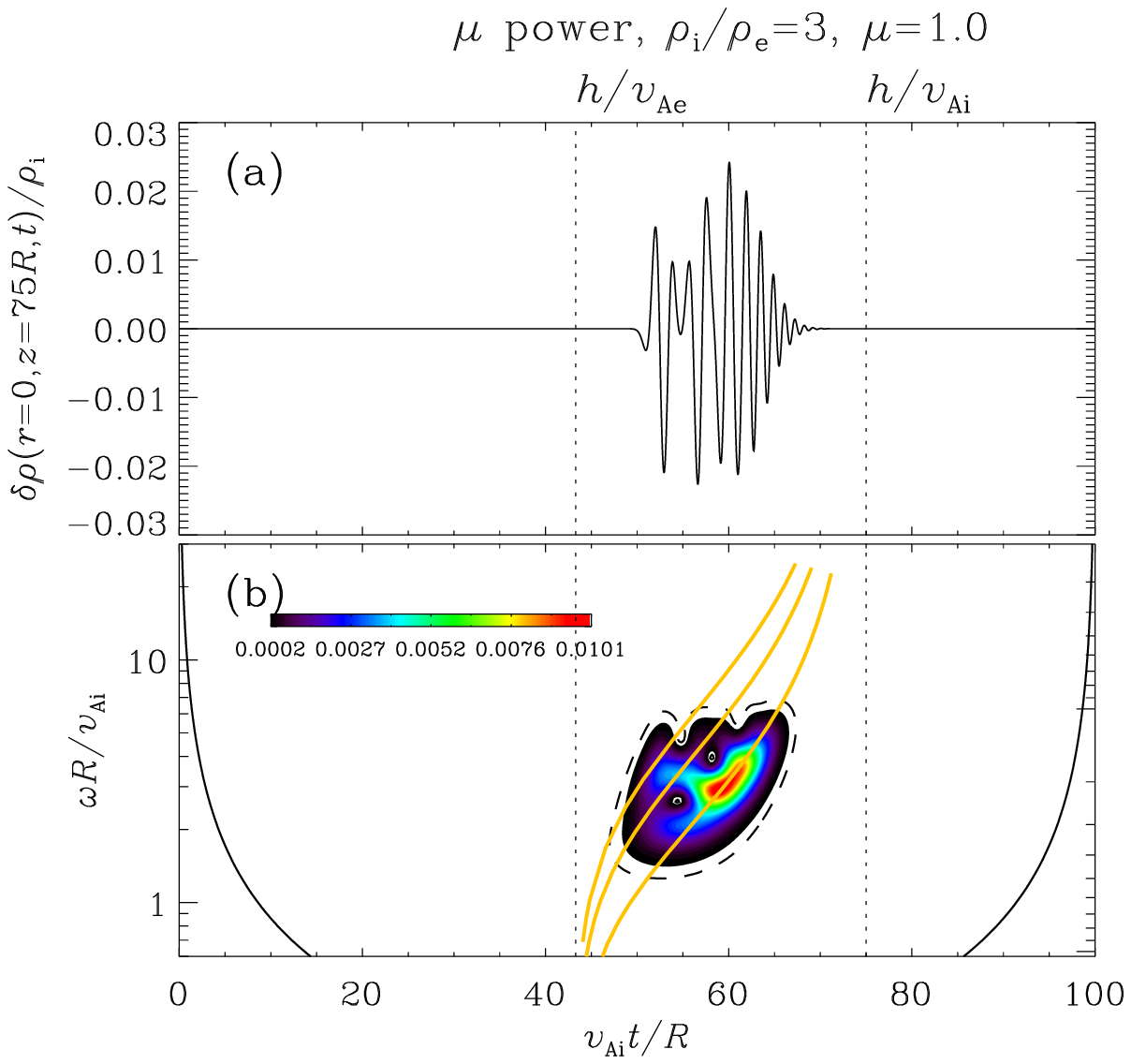}
\includegraphics[height=80mm]{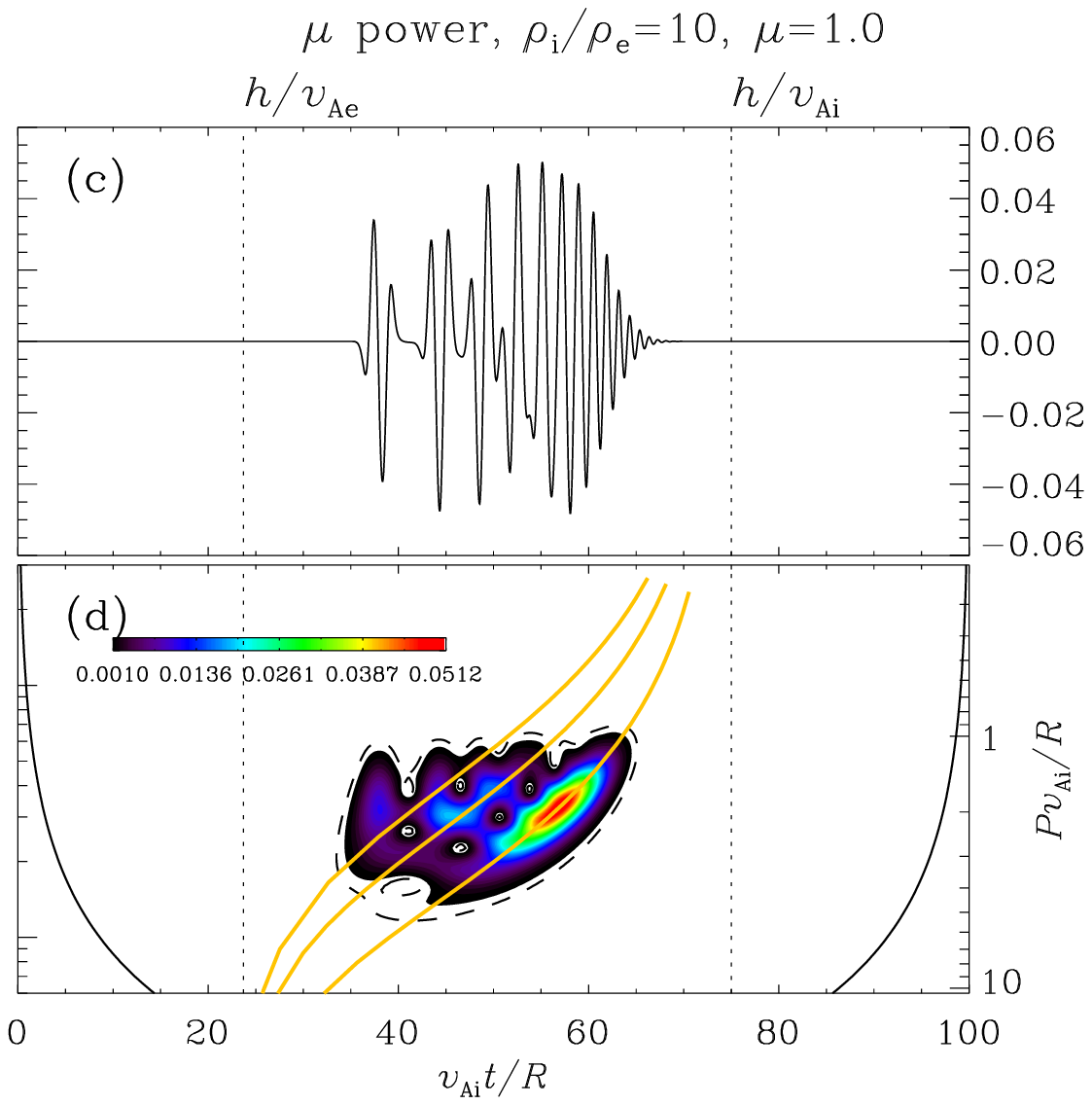}
\caption{
 Similar to Fig.~\ref{fig_wavelet_tophat} but for {``$\mu$ power''} profiles with $\mu=1$.
}
\label{fig_wavelet_monomu1}
\end{figure}

\begin{figure}
\centering
\includegraphics[height=80mm]{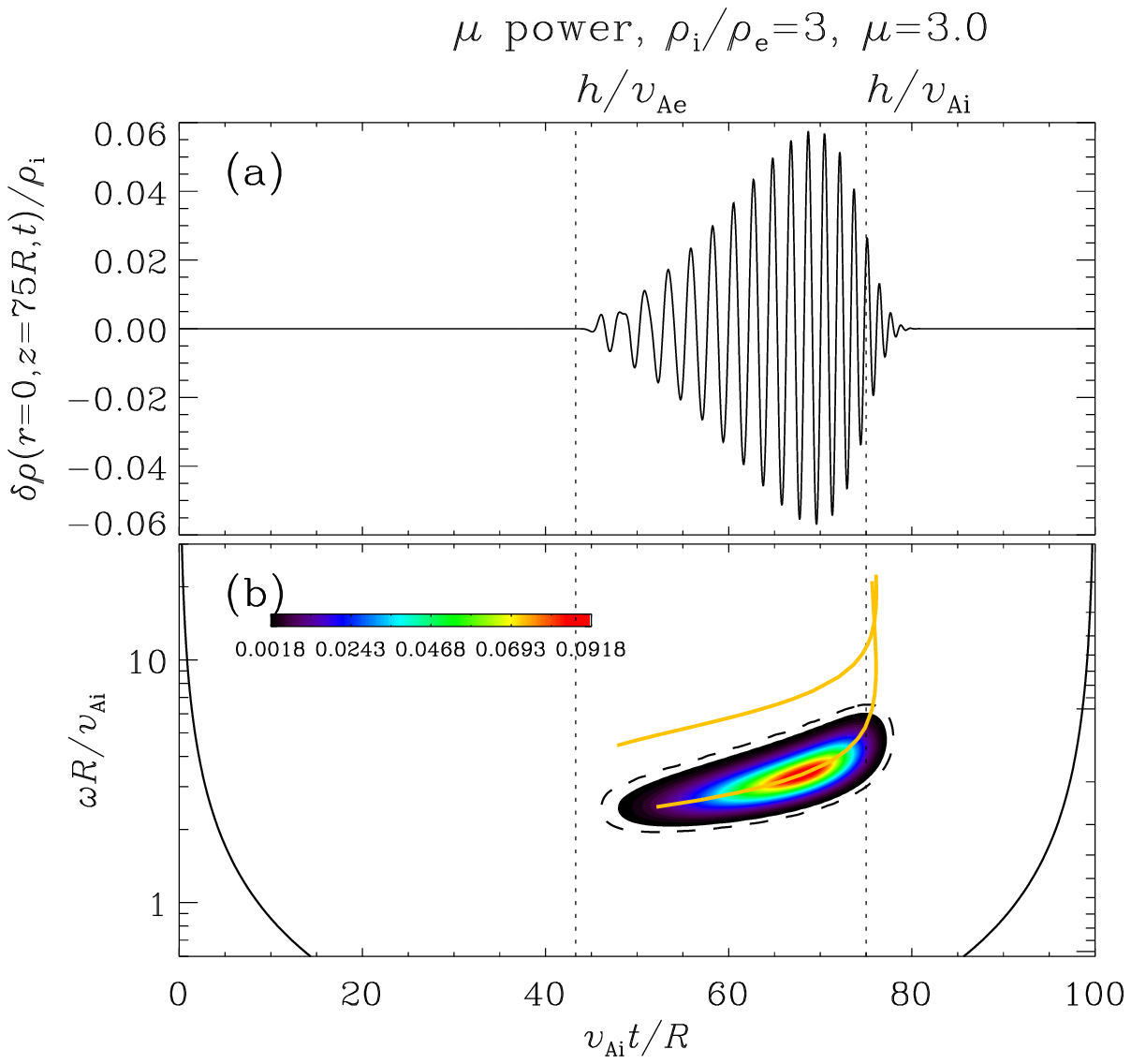}
\includegraphics[height=80mm]{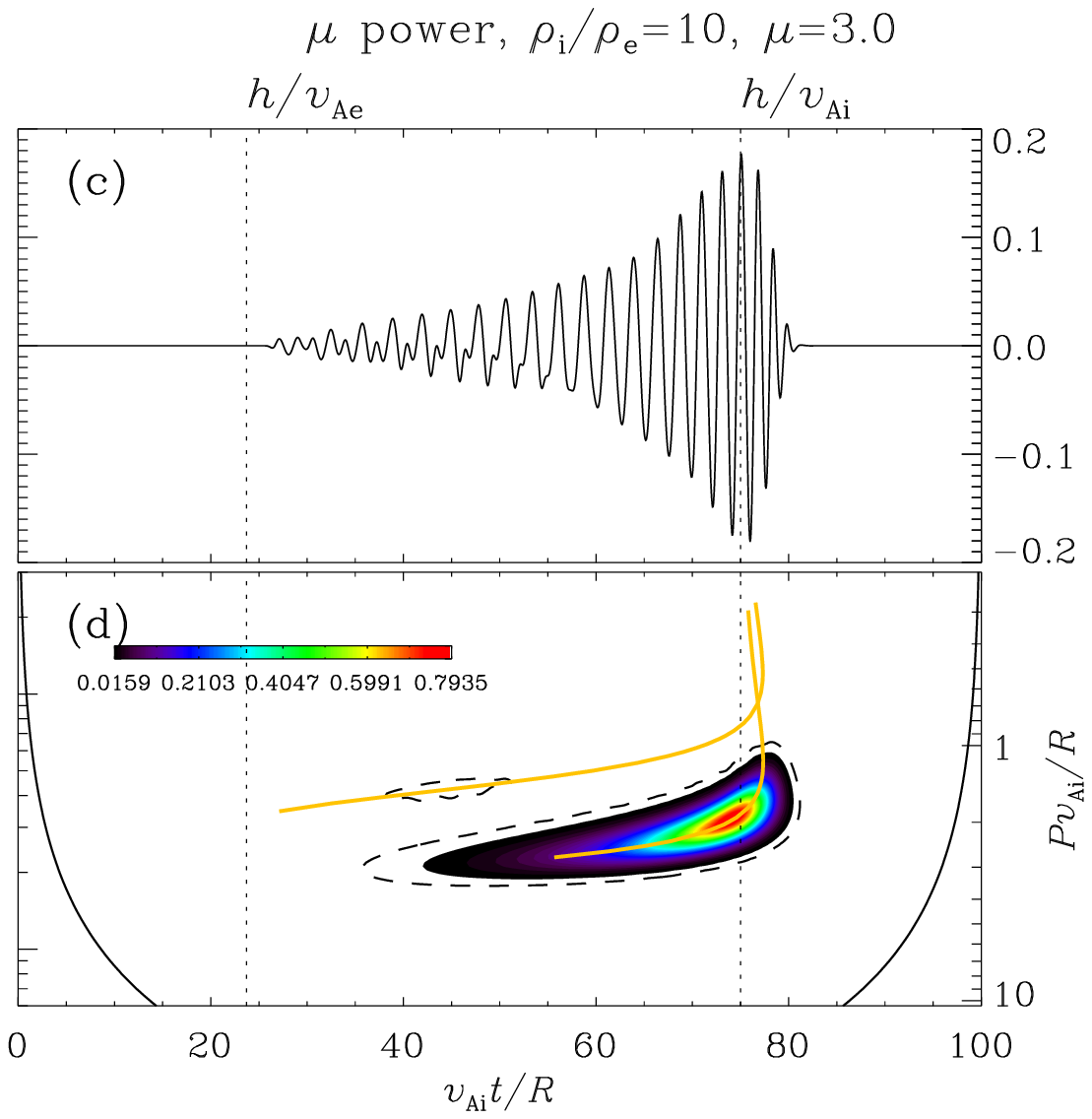}
\caption{
 Similar to Fig.~\ref{fig_wavelet_tophat} but for {``$\mu$ power''} profiles with $\mu=3$.
}
\label{fig_wavelet_monomu3}
\end{figure}

\begin{figure}
\centering
\includegraphics[width=0.9\columnwidth]{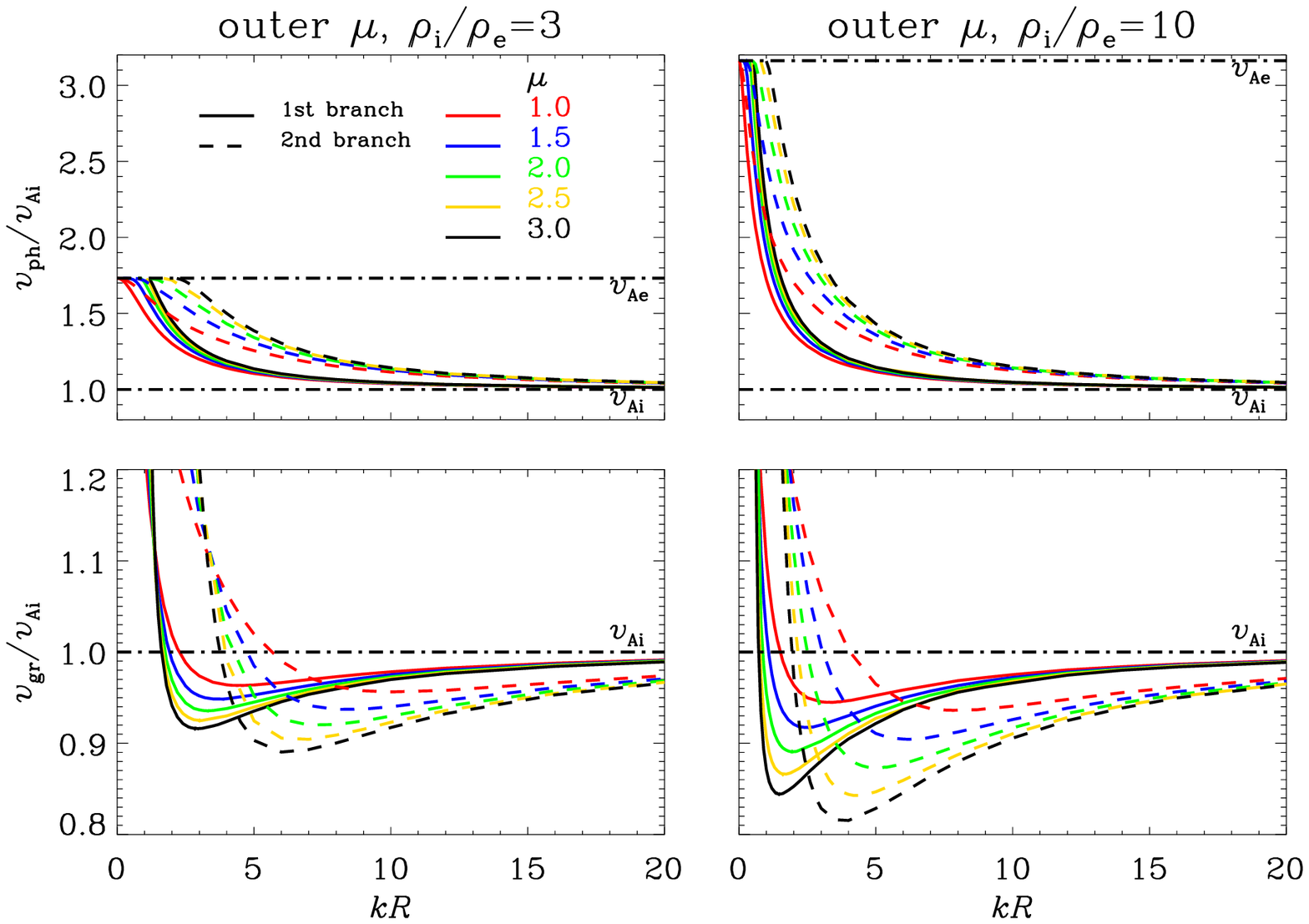}
 \caption{
 Similar to Fig.~\ref{fig_vphvg_k_tophat} but for ``outer $\mu$'' profiles  with a number of $\mu$ as labeled.
}
 \label{fig_vphvg_k_outermu}
\end{figure}

\begin{figure}
\centering
\includegraphics[height=80mm]{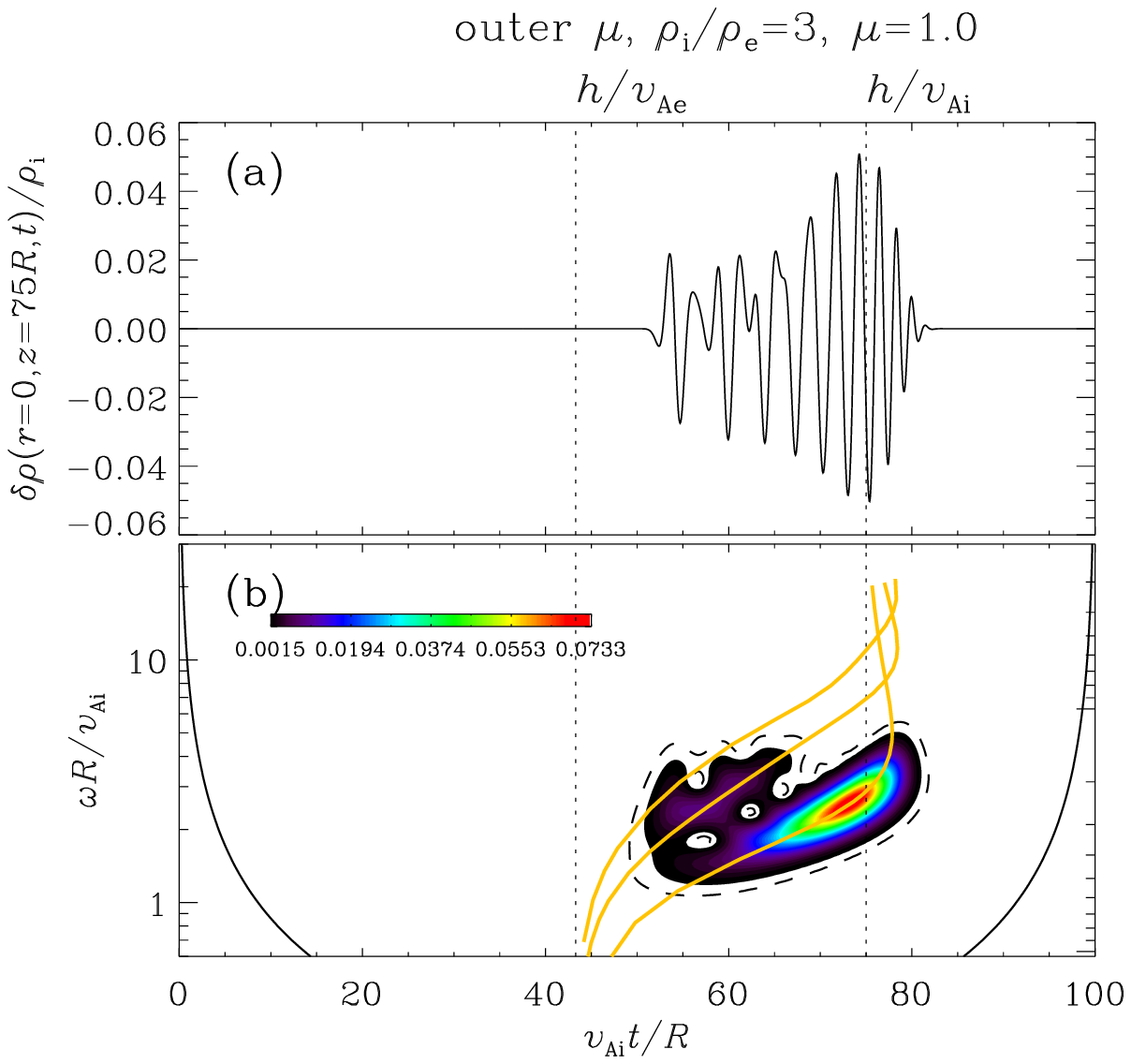}
\includegraphics[height=80mm]{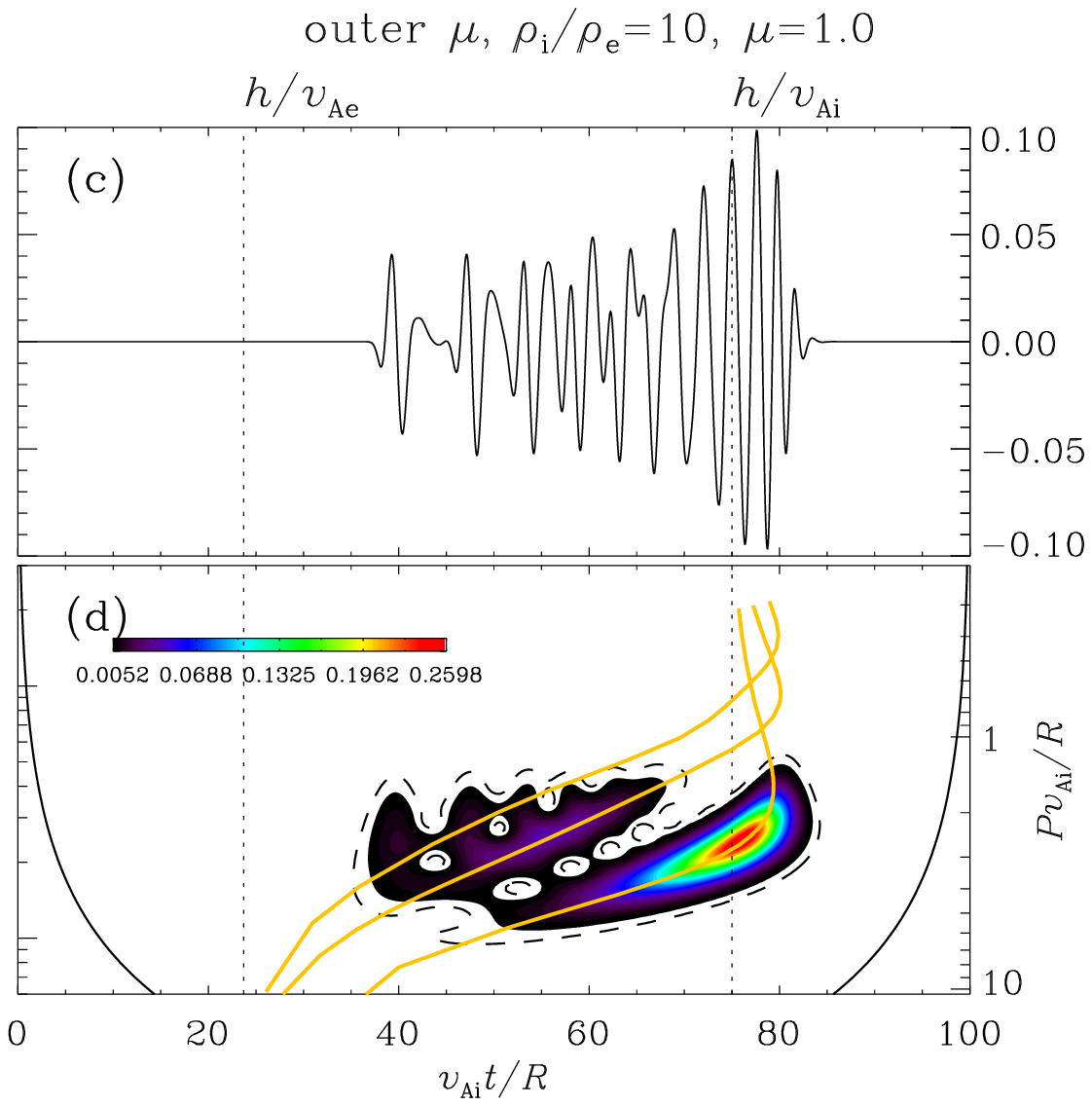}
\caption{
 Similar to Fig.~\ref{fig_wavelet_tophat} but for ``outer $\mu$'' profiles with $\mu=1$.
}
\label{fig_wavelet_outermu1}
\end{figure}

\begin{figure}
\centering
\includegraphics[height=80mm]{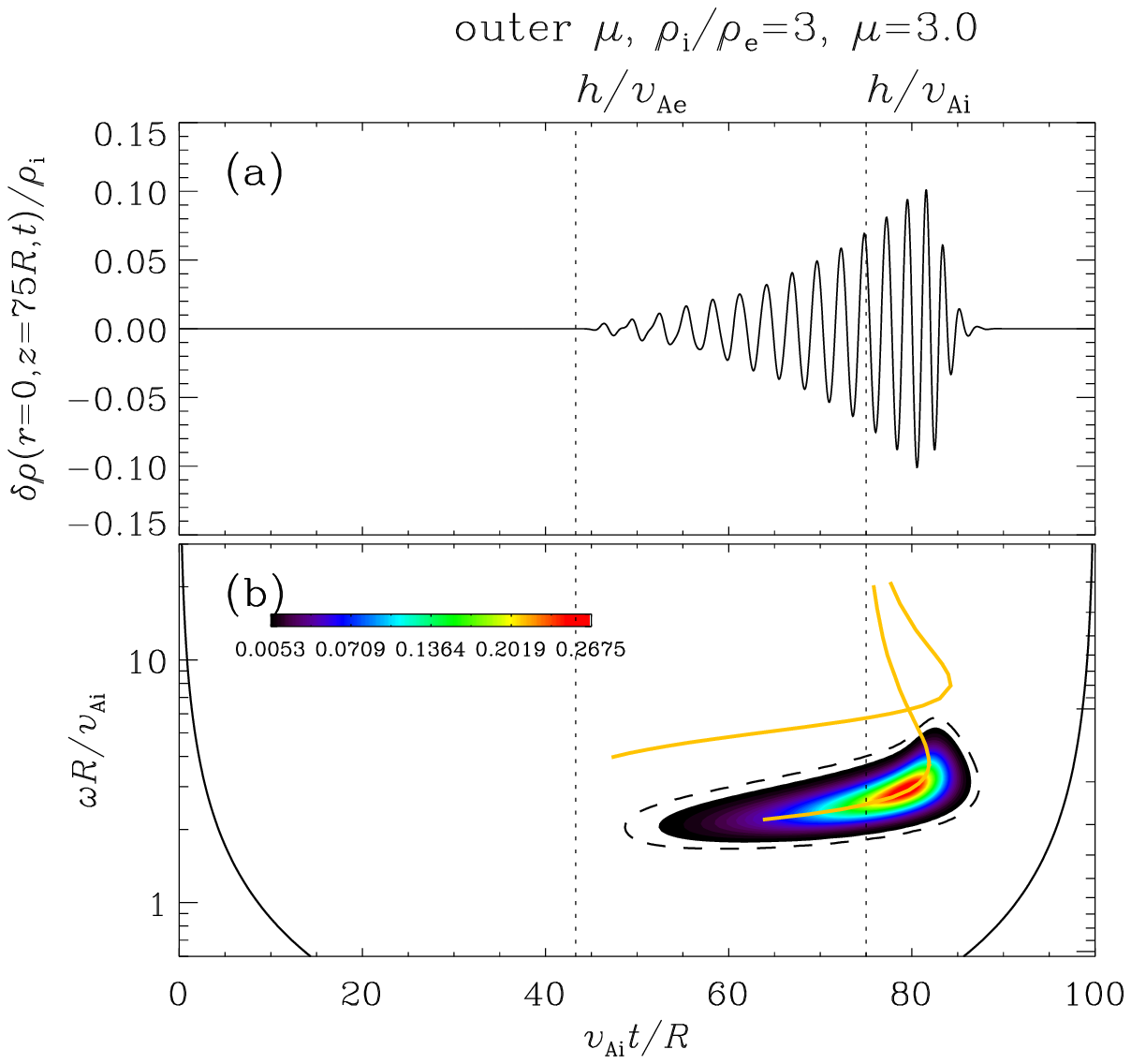}
\includegraphics[height=80mm]{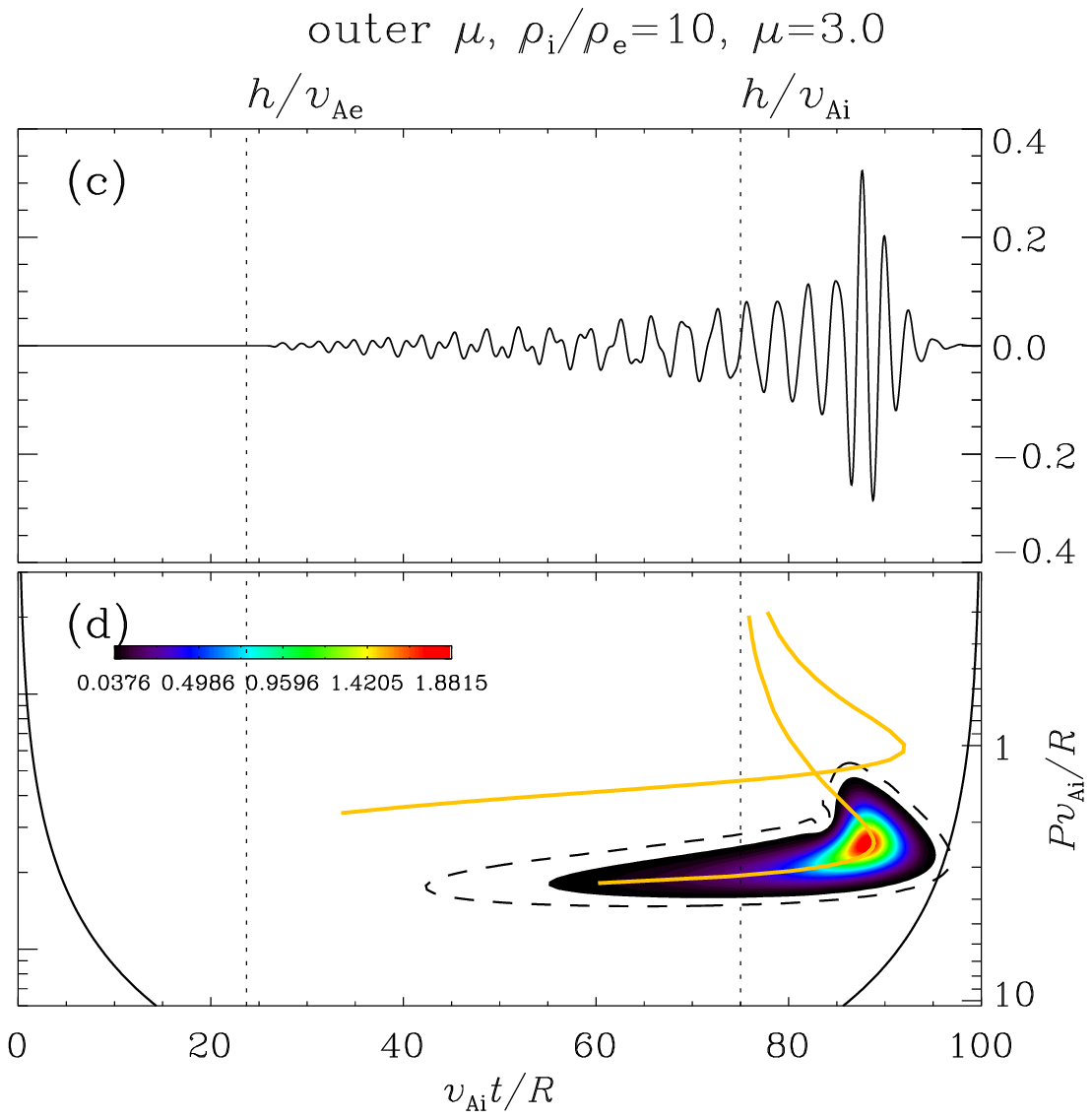}
\caption{
 Similar to Fig.~\ref{fig_wavelet_tophat} but for ``outer $\mu$'' profiles with $\mu=3$.
}
\label{fig_wavelet_outermu3}
\end{figure}

\begin{figure}
\centering
\includegraphics[width=0.9\columnwidth]{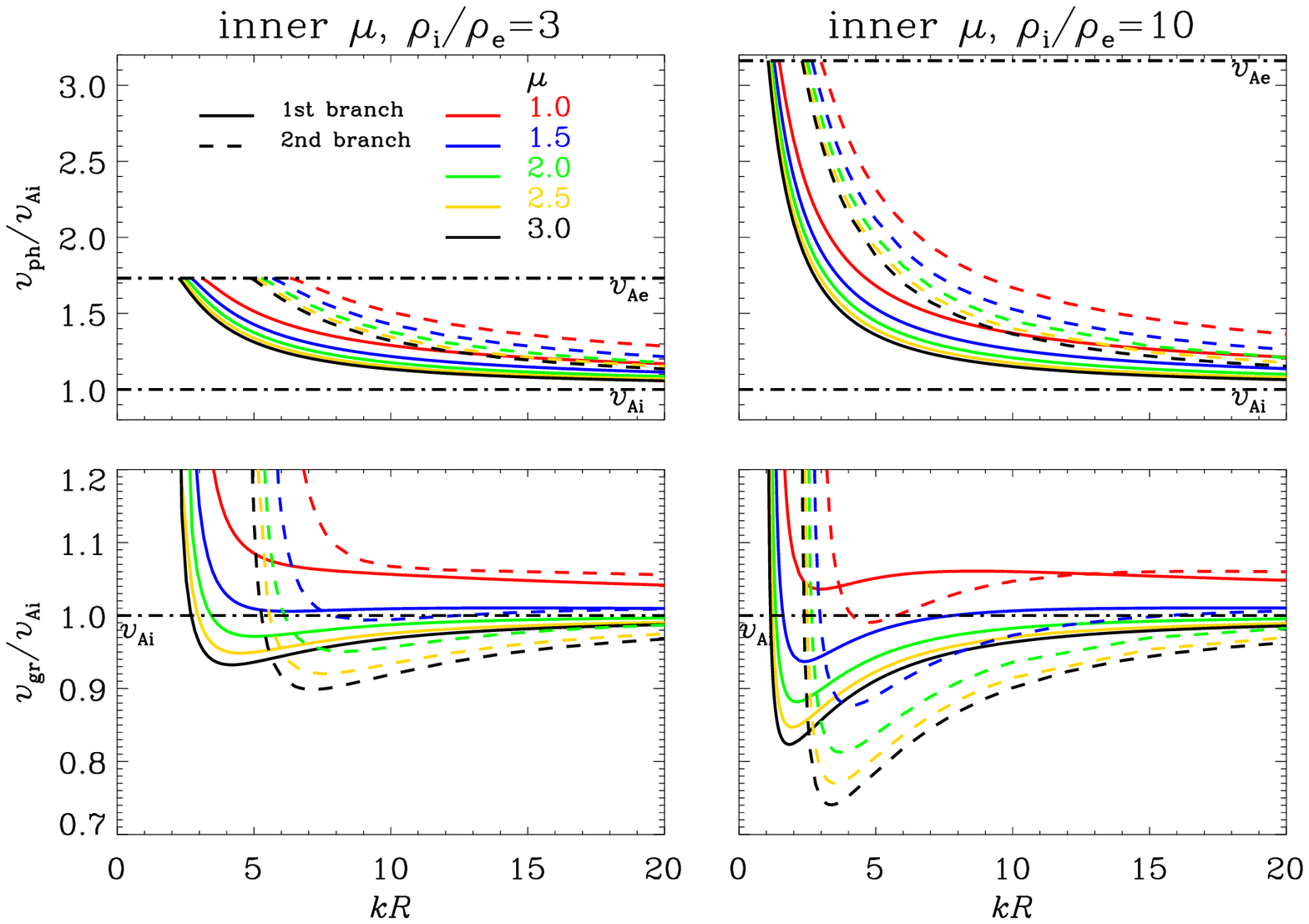}
 \caption{
 Similar to Fig.~\ref{fig_vphvg_k_tophat} but for ``inner $\mu$'' profiles  with a number of $\mu$ as labeled.
}
 \label{fig_vphvg_k_innermu}
\end{figure}

\begin{figure}
\centering
\includegraphics[width=0.6\columnwidth]{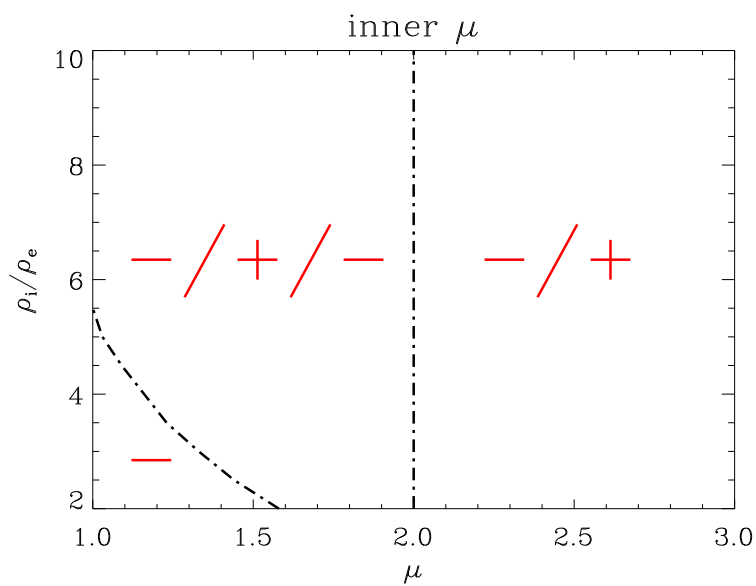}
 \caption{
{Summary of the properties of the $\vgr-k$ curves for ``inner $\mu$'' profiles.
 The symbol ``$-$'' represents the tendency for $\vgr$ to decrease monotonically with $k$,
     while ``$-/+$'' corresponds to the cases where $\vgr$ decreases with $k$ first before
     increasing again.
 Furthermore, ``$-/+/-$'' represents the cases where $\vgr$ decreases first and then increases
     before decreasing towards $\vai$.}
}
 \label{fig_DRsum_innermu}
\end{figure}

\begin{figure}
\centering
\includegraphics[height=80mm]{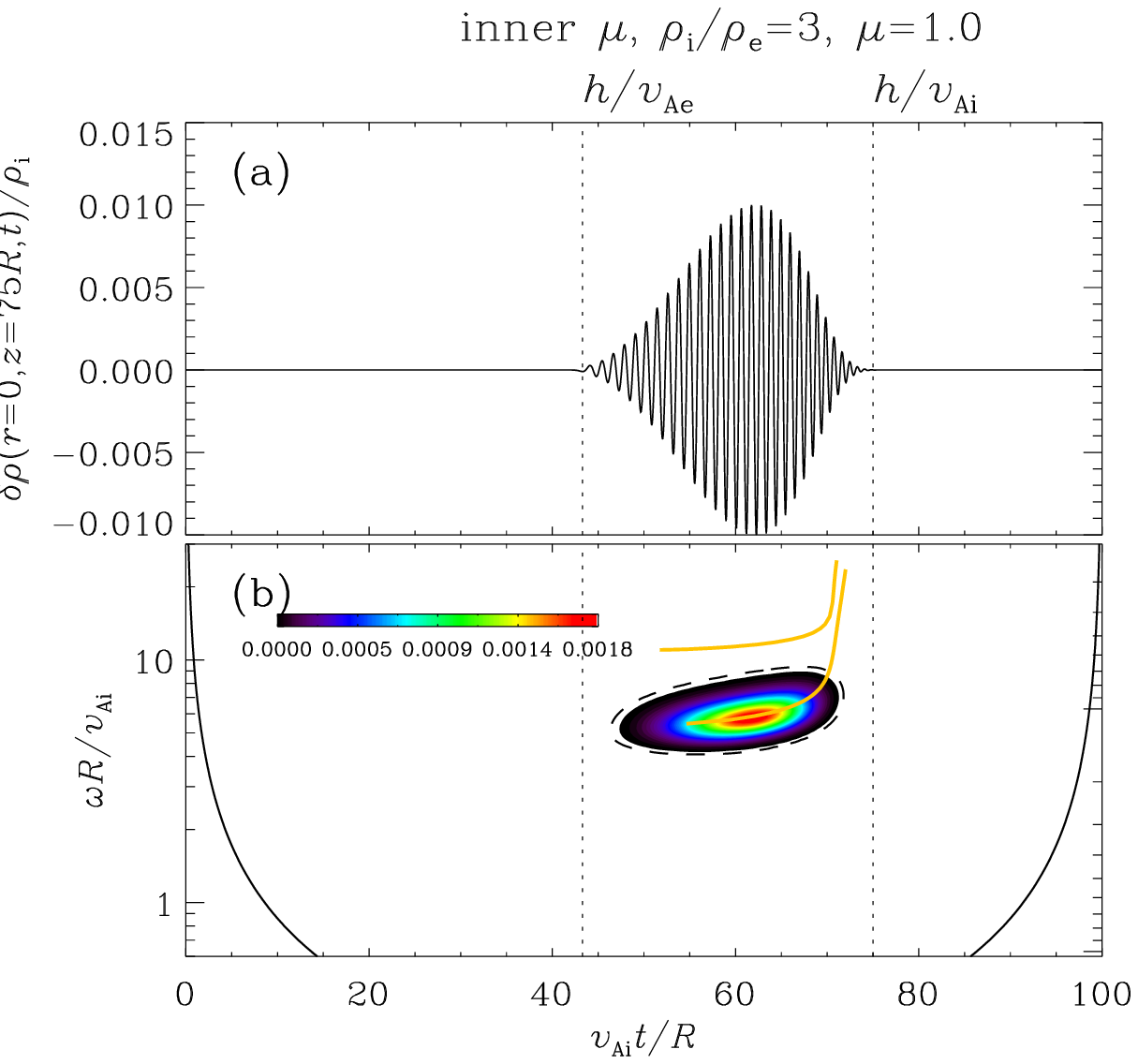}
\includegraphics[height=80mm]{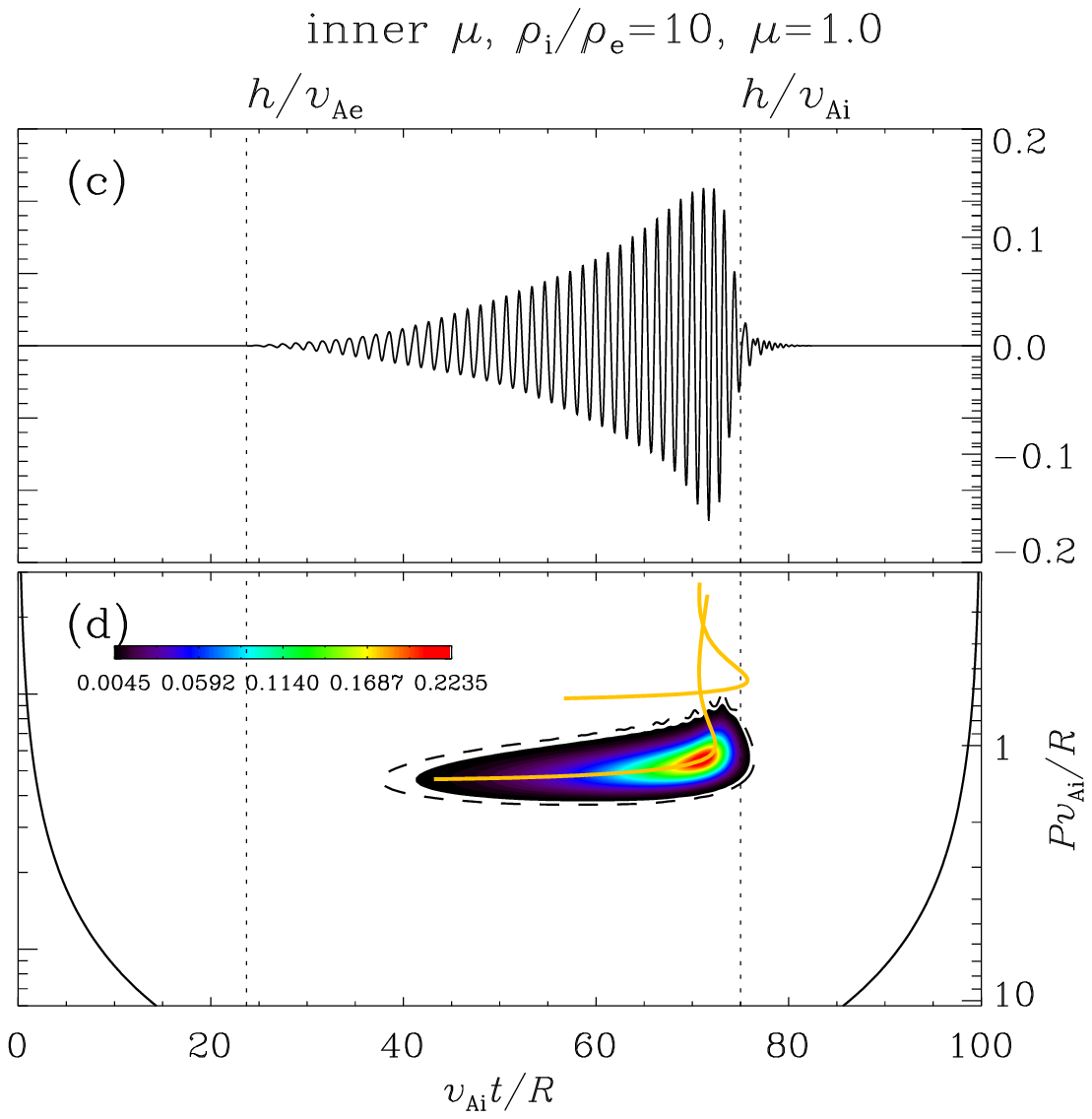}
\caption{
 Similar to Fig.~\ref{fig_wavelet_tophat} but for ``inner $\mu$'' profiles with $\mu=1$.
}
\label{fig_wavelet_innermu1}
\end{figure}

\begin{figure}
\centering
\includegraphics[height=80mm]{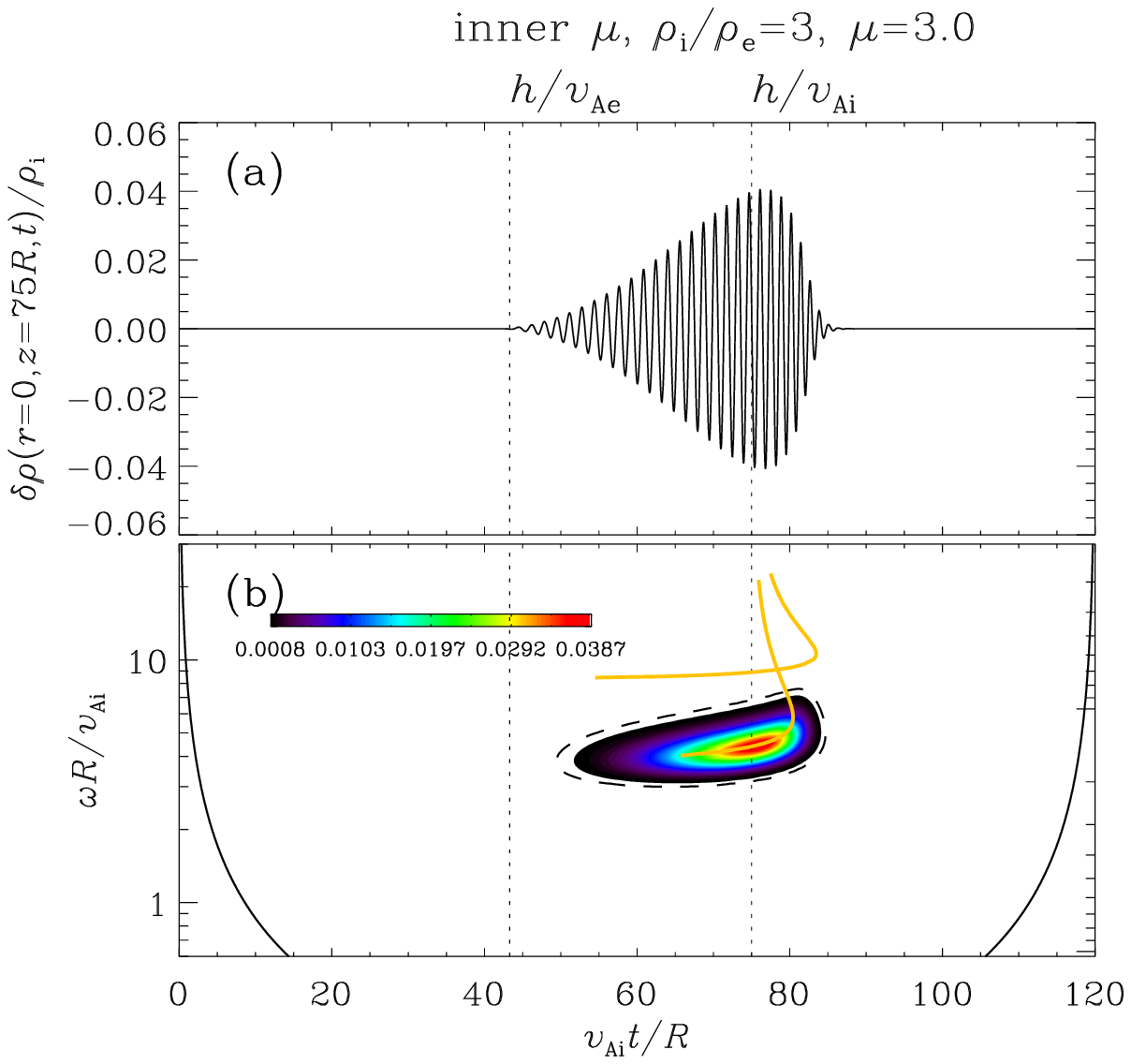}
\includegraphics[height=80mm]{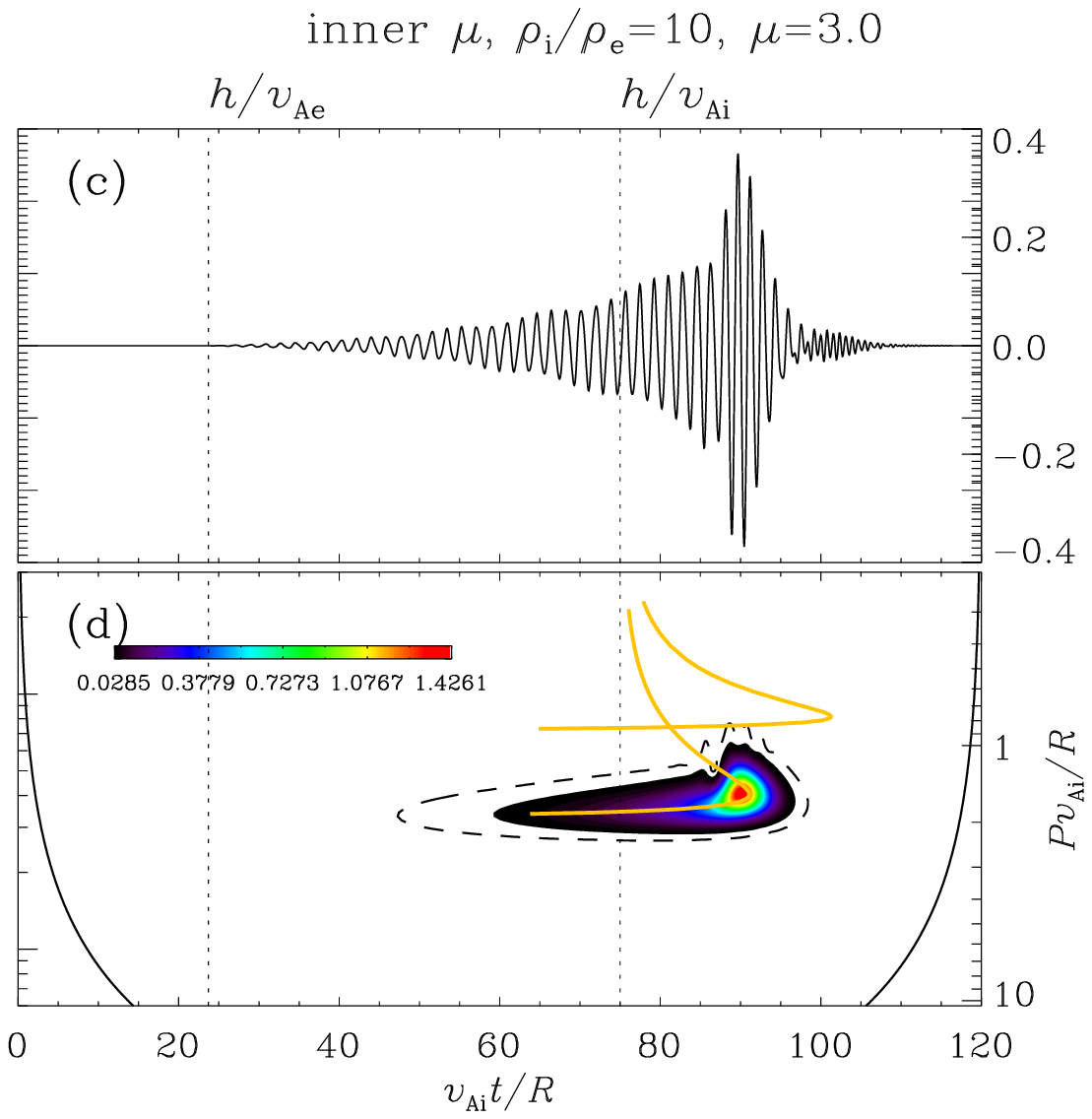}
\caption{
 Similar to Fig.~\ref{fig_wavelet_tophat} but for ``inner $\mu$'' profiles with $\mu=3$.
}
\label{fig_wavelet_innermu3}
\end{figure}

{Note that the temporal and wavelet signatures presented here pertain to the perturbations $\delta \rho$ sampled at
    a distance $h$ sufficiently far from an impulsive driver, for which we choose a fixed combination
    of $[\sigma_r, \sigma_z]$, the parameters
    characterizing the spatial extent to which the initial perturbation spans
    in the radial and axial directions.
Some further computations indicate that the temporal and wavelet signatures also depend on where $\delta\rho$ is sampled,
    as well as on how the values of $\sigma_r$ and $\sigma_z$ are specified. 
Instead of adding these further results to this already long paper, let us offer a discussion in the framework proposed 
    by \citeauthor{2015ApJ...806...56O}~(\citeyear{2015ApJ...806...56O},  
    see also \citeauthor{2014ApJ...789...48O}~\citeyear{2014ApJ...789...48O} where kink waves are addressed).
Theoretically speaking, the temporal evolution of impulsively generated wave trains is determined by three physical ingredients.
First, the properties of both proper and improper modes.
These are determined by the parameters characterizing the equilibrium, namely $[f(r), \rhoi/\rhoe]$ with our notations.
Note that proper modes translate into trapped modes, and improper modes are intimately related to leaky modes. 
Second, the spatial extent of the initial perturbations, denoted by $[\sigma_r, \sigma_z]$ in our study. 
This determines 
    how the energy contained in the initial perturbations is apportioned between proper
    and improper modes in general, and how it is further distributed 
    into different wavelength or frequency ranges. 
Third, the fact that the contribution to the wave trains from improper modes is different from that due to proper modes
    given their different attenuation rates.
By choosing a value for $h$ to be as large as $75 R$, we made sure that only
    the components pertinent to proper (or trapped) modes survive.
On the other hand, choosing a single combination $[\sigma_r, \sigma_z]$ was meant to highlight only  
    the importance of transverse structuring in determining the properties of trapped modes.
Adopting the same $h$ but experimenting with other choices of $[\sigma_r, \sigma_z]$, we find that
    while the wavelet spectra are still organized by the $\omega-h/\vgr$ curves, 
    their specific distributions can look different.    
Take top-hat profiles for simplicity, and consider what happens when only $\sigma_z$ is varied by 
    choosing $\sigma_z = \sqrt{2} R$ (twice the nominal value in the text).
Evidently, one expects from the description of the initial perturbation (Eq.~\ref{eq_IC_vr})
    that more energy will be partitioned to lower wavenumbers.
As a consequence, while the Morlet spectrum for $\rhoi/\rhoe=10$ 
    still looks like a ``crazy tadpole'', its head is not as broad as 
    in Fig.~\ref{fig_wavelet_tophat}d found for $\sigma_z = R/\sqrt{2}$.
For a density contrast of $3$, the shift of the Morlet power towards 
    the low-frequency portion reaches such a point that the broad head disappears altogether.
In fact, the consequence of varying $\sigma_z$ on impulsively generated wave trains
    was already clear in \citet{2014ApJ...789...48O} and \citet{2015ApJ...806...56O}.
For instance, the disappearance of the short-wavelength components can be readily discerned
    in Fig.~8a of \citet{2014ApJ...789...48O}  when compared with Fig.~7a therein.
}    
    
{
The possible effects due to $[\sigma_r, \sigma_z]$ aside}, 
    our choices of profile descriptions cannot exhaust
    the largely unknown density distributions transverse to coronal tubes.
{One may question the validity of the classification of the Morlet spectra into the above-mentioned categories,
    if adopting radial density profiles other than employed here.
For this purpose, we may consider the profiles examined in our paper I, where 
    the density distribution comprises a uniform cord, a uniform external medium,
    and a transition layer sandwiched between.
The $\omega-k$ curves all possess cutoff wavenumbers given that the density profile outside the tube 
    falls with distance more rapidly than $r^{-2}$.
In this regard, they belong to either category I or IV, and the corresponding Morlet spectra
    largely agree with the typical features obtained in the present study
    (see Figs. 10 to 15 in paper I).
We need to mention that strictly speaking, Fig.~13 therein pertains to category I but the Morlet spectrum
    seems typical of category IV.
As explained there, this takes place because the frequency at which the group speed minimum is attained is so high
    that essentially no energy is received in that frequency range.
Consequently, the Morlet spectrum behaves as if the group speed minimum were absent, meaning that 
    the $\omega-k$ curves are effectively monotonical.
This once again strengthens the importance of the spatial distribution of the energy contained 
    in the initial perturbations.}

What our computations indicate is that the behavior of the $\vgr-k$ curves
    can be quite rich, and this rich variety can indeed be reflected
    in the temporal evolution and Morlet spectra of impulsively generated wave trains.
Given that the characteristics of the $\vgr-k$ curves depend on how density profiles are described,
    it then seems imperative to dig into the available high-cadence data to look for
    those Morlet spectra that look {drastically} different from ``crazy tadpoles''.
{Let us stress here that the behavior of the $\vgr-k$ curves is not the only cause for 
    the Morlet spectra other than ``crazy tadpoles'' to arise.
A more complete picture can be reached by further considering the spatial extent of the initial perturbations,
    which themselves are also largely unknown.
A study along this line of thinking is underway, but is beyond the scope of the present manuscript though.    
}

\acknowledgments
{We thank the referee for the constructive comments,
     which helped improve this manuscript substantially.}
This work is supported by
    the National Natural Science Foundation of China (BL:41674172, 41474149, 41174154, and 41274176; SXC:41604145, MX:41374175),
    and by the Provincial Natural Science Foundation of Shandong via Grants JQ201212 (BL) and ZR2016DP03 (HY).

\bibliographystyle{apj}
\bibliography{impls}

\IfFileExists{\jobname.bbl}{} {\typeout{}
\typeout{****************************************************}
\typeout{****************************************************}
\typeout{** Please run "bibtex \jobname" to obtain} \typeout{**
the bibliography and then re-run LaTeX} \typeout{** twice to fix
the references !}
\typeout{****************************************************}
\typeout{****************************************************}
\typeout{}}

\clearpage

\clearpage

\clearpage

\clearpage

\clearpage

\clearpage

\clearpage

\clearpage

\clearpage

\clearpage

\clearpage

\clearpage

\clearpage

\end{document}